\documentstyle[aps,floats,epsf]{revtex}
\begin{document}                
\def\be{\begin{equation}}
\def\ee{\end{equation}}
\def\bea{\begin{eqnarray}}
\def\eea{\end{eqnarray}}
\setlength{\epsfxsize}{4.0in}\title{PAIRED ACCELERATED FRAMES: 
THE PERFECT INTERFEROMETER WITH EVERYWHERE SMOOTH WAVE 
AMPLITUDES\footnote{Published in Phys. Rev. D {\bf 59}, 104009 (1999) and
based in part on a talk given at the Sixth Midwestern 
Relativity Conference, St. Louis, November 13-14,1997, and 
on part of a report given in the session ``Casimir Effect'' 
originally published in the Proc. of the Eighth Marcel Grossmann
Meeting on General Relativity, Jerusalem, June 23-27, 1997, edited by
Tsvi Piran (World Scientific, Singapore, 1999)  }}

\author{ULRICH H. GERLACH}
\address{Department of Mathematics, Ohio State University, Columbus, OH
43210, USA}
%
\maketitle
\begin{abstract}                
  
  In the absence of gravitation the distinguishing feature of any
  linearly and uniformly accelerated frame is that it is one member of
  a pair moving into opposite directions. This pairing partitions
  Minkowski spacetime into four mutually exclusive and jointly
  exhaustive domains whose boundary consists of the future and past
  event horizons relative to each of the two frames. This
  acceleration-induced partitioning of spacetime leads to a
  nature-given interferometer. It accommodates quantum mechanical and
  wave mechanical processes in spacetime which in (Euclidean) optics
  correspond to wave processes in a ``Mach-Zehnder'' interferometer:
  amplitude splitting, reflection, and interference. The spacetime
  description of these processes is given in terms of amplitudes which
  are defined globally as well as locally in each of the four Rindler
  spacetime domains. It is found that these amplitudes behave smoothly
  across the event horizons. In this context there arises quite
  naturally a complete set of orthonormal wave packet histories, one
  of whose key properties is their \emph{explosivity index}. It
  measures the rate at which the wave packets collapse and re-explode.
  In the limit of low index values the wave packets trace out fuzzy
  world lines. For large mass this fuzziness turns into the
  razor-sharpness of classically determinate world lines of the
  familiar Klein-Gordon scalar particles. By contrast, in the
  asymptotic limit of high index values, there are no world lines, not
  even fuzzy ones.  Instead, the wave packet histories are those of
  entities with non-trivial internal collapse and explosion dynamics.
  Their details are described by the wave processes in the
  above-mentioned Mach-Zehnder interferometer. Each one of them is a
  double slit interference process. These wave processes are applied
  to elucidate the amplification of waves in an accelerated
  inhomogeneous dielectric. Also discussed are the properties and
  relationships among the transition amplitudes of an accelerated
  finite-time detector.

\noindent PACS numbers:  04.62.+v, 03.65.Pm, 42.25.Hz, 42.25.Fx

\end{abstract}
\section{INTRODUCTION}
The fundamental feature of uniformly and linearly accelerated observers
is that they come in pairs. If the spacetime of one accelerated frame is
\begin{equation}
\left. 
\begin{array}{c}
  t-t_0 =\xi\sinh \tau \\
  z-z_0  =\xi  \cosh \tau
  \end{array} 
\right\} \quad 
  \hbox{``Rindler Sector I''} \quad ,
\label{eq:Rindler I}
\end{equation}
then the spacetime of its twin is 
\begin{equation}
\left. 
\begin{array}{c}
  t-t_0 =-\xi\sinh \tau \\
  z-z_0  =-\xi  \cosh \tau
  \end{array} 
\right\} \quad 
  \hbox{``Rindler Sector II''} \quad .
\label{eq:Rindler II}
\end{equation}
These two regions of spacetime are causally disjoint. Their event horizons
$ t-t_0 =\pm \vert z-z_0 \vert$ separate these regions 
from their chronological 
future $F$,
\begin{equation}
\left. 
\begin{array}{c}
  t-t_0 =\xi\cosh \tau \\
  z-z_0  =\xi  \sinh \tau
  \end{array} 
\right\} \quad 
  \hbox{``Rindler Sector F''} \quad ,
\label{eq:Rindler F}
\end{equation}
and their chronological past $P$,
\begin{equation}
\left. 
\begin{array}{c}
  t-t_0 =-\xi\cosh \tau \\
  z-z_0  =-\xi  \sinh \tau
  \end{array} 
\right\} \quad 
  \hbox{``Rindler Sector P''} \quad .
\label{eq:Rindler P}
\end{equation}
We always assume that $\xi >0$. Thus, as shown in Figure 1, an 
accelerated observer induces a
partitioning of Minkowski spacetime into the four Rindler sectors
$I,II,F$ and $P$.
\begin{figure}[h!]
\epsfclipon
  \setlength{\epsfysize}{3.5in}
  \setlength{\epsfxsize}{6.0in}
\centerline{\epsffile[0 400 500 680]{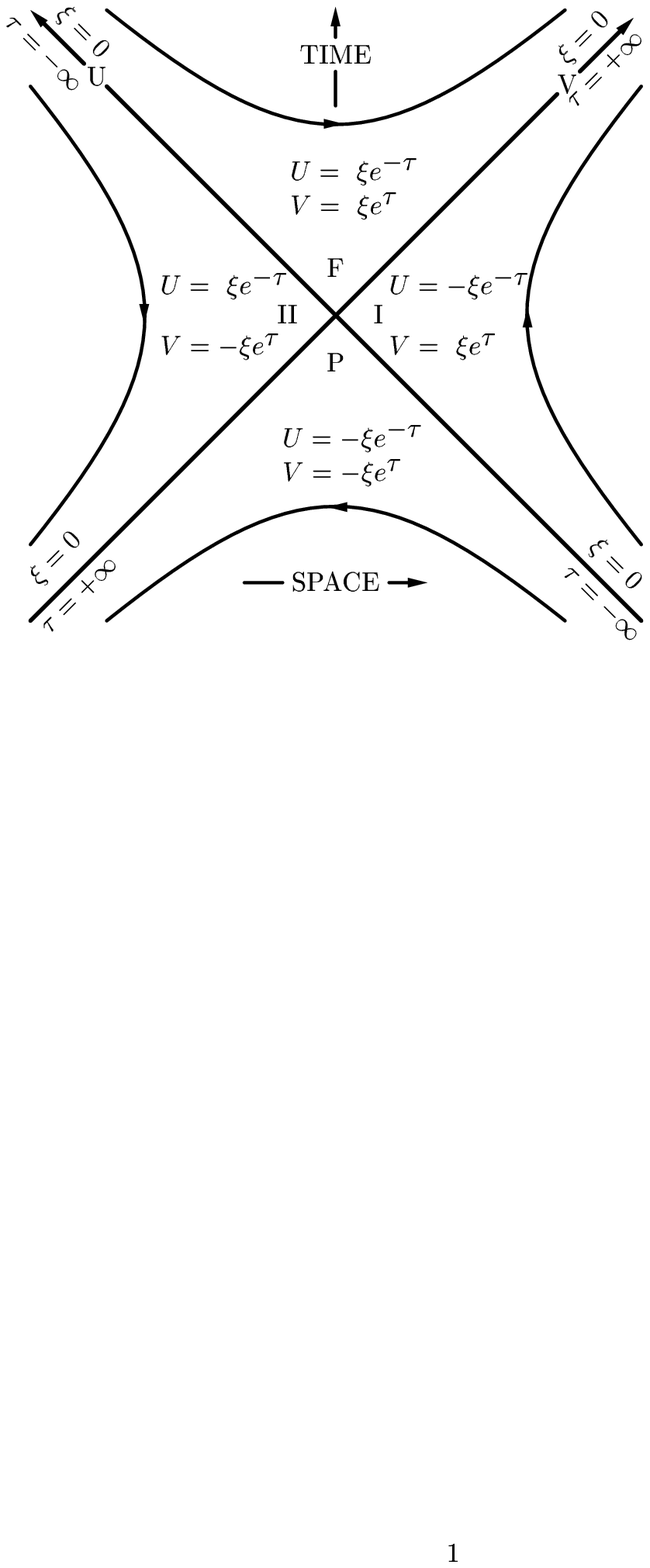}}
       \epsfverbosetrue
\caption{ Acceleration-induced partitioning of spacetime into the
four Rindler sectors.}
\label{fig:Rindler spacetime}
\end{figure}

This partitioning opens a new perspective for relativistic processes 
governed by the Klein-Gordon wave equation: {\it The four Rindler sectors
form a perfect interferometer.}

\subsection{Lorentzian Mach-Zehnder Interferometer}

\begin{figure}
\epsfclipon

\epsffile[100 570 500 680]{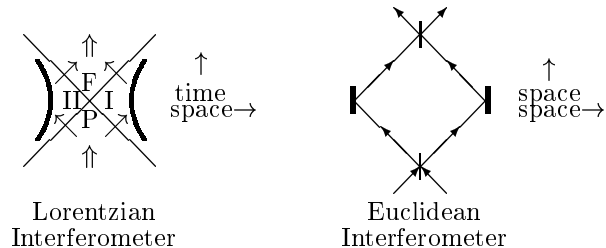}
       \epsfverbosetrue
\caption{ Two interferometers. The Euclidean interferometer, which
in optics is known as a Mach-Zehnder interferometer, consists of
a half-silvered entrance mirror (thin line at the bottom),
a half-silvered exit mirror (thin line at the top),
and two fully reflecting mirrors (thick lines on the right and the left).
The Lorentzian interferometer consists of the entrance region, Rindler
Sector $P$,
the exit region, Rindler Sector $F$, and the two reflective regions,
Rindler Sectors $I$ and $II$. The reflection is brought about by 
by the pseudo-gravitational potential (see Eq. (\ref{eq:Rindler equation}))
in each sector. The thick hyperbolas
separate the region of this potential where the mode is oscillatory 
from that region where 
it is evanescent. This reflection process is detailed in 
Figure~\ref{fig:reflected modes}.}
\label{fig:two interferometers}
\end{figure}

More precisely, the two accelerating frames ($I\& II$) serve as the
two coherent legs of a Lorentzian version of the Mach-Zehnder
interferometer\cite{Born and Wolf}. As indicated in 
Figure~\ref{fig:two interferometers}, 
\smallskip
the two well-known pseudo-gravitational potentials in these frames
serve as the two mirrors with 100\% reflectivity. The two spacetime
regions in the future $F$ and the past $P$ \emph{near} the
intersection (``bifurcation event'') of the event horizons serve as
``half-silvered'' mirrors. A wave in $P$ \emph{far} from the
bifurcation event enters the ``interferometer'' from $P$.  One can
show mathematically, and this is done in
Figures~\ref{fig:amplitude splitting} and \ref{fig:WKB modes}, 
that near the two event horizons
$t-t_0=-\vert x-x_0\vert$ the wave splits into two partial waves: one
propagates across the past event horizon, enters Rindler Sector
$I$. There it gets reflected by the Lorentz invariant
pseudo-gravitational potential. This reflection is demonstrated
mathematically in Figures \ref{fig:reflected modes} and \ref{fig:WKB modes}. 
The other partial
wave suffers the analogous reflection in Rindler Sector $II$.  The two
partial waves cross the event horizons $t-t_0=\vert x-x_0\vert$ and,
as shown in Figure \ref{fig:amplitude splitting}, recombine in Rindler Sector
$F$. There these waves are intercepted by two freely floating
detectors configured to perform a time-plus-space version of
Wheeler's ``delayed choice''
experiment~\cite{Wheeler}. This means that an (inertial) physicist who
spends his life in Rindler Sector $F$

\noindent (a) releases two free-float particle detectors whose apertures
point into opposite directions.

\begin{figure}
\epsfclipon
\epsfxsize=6in
\epsffile[100 420 612 670]{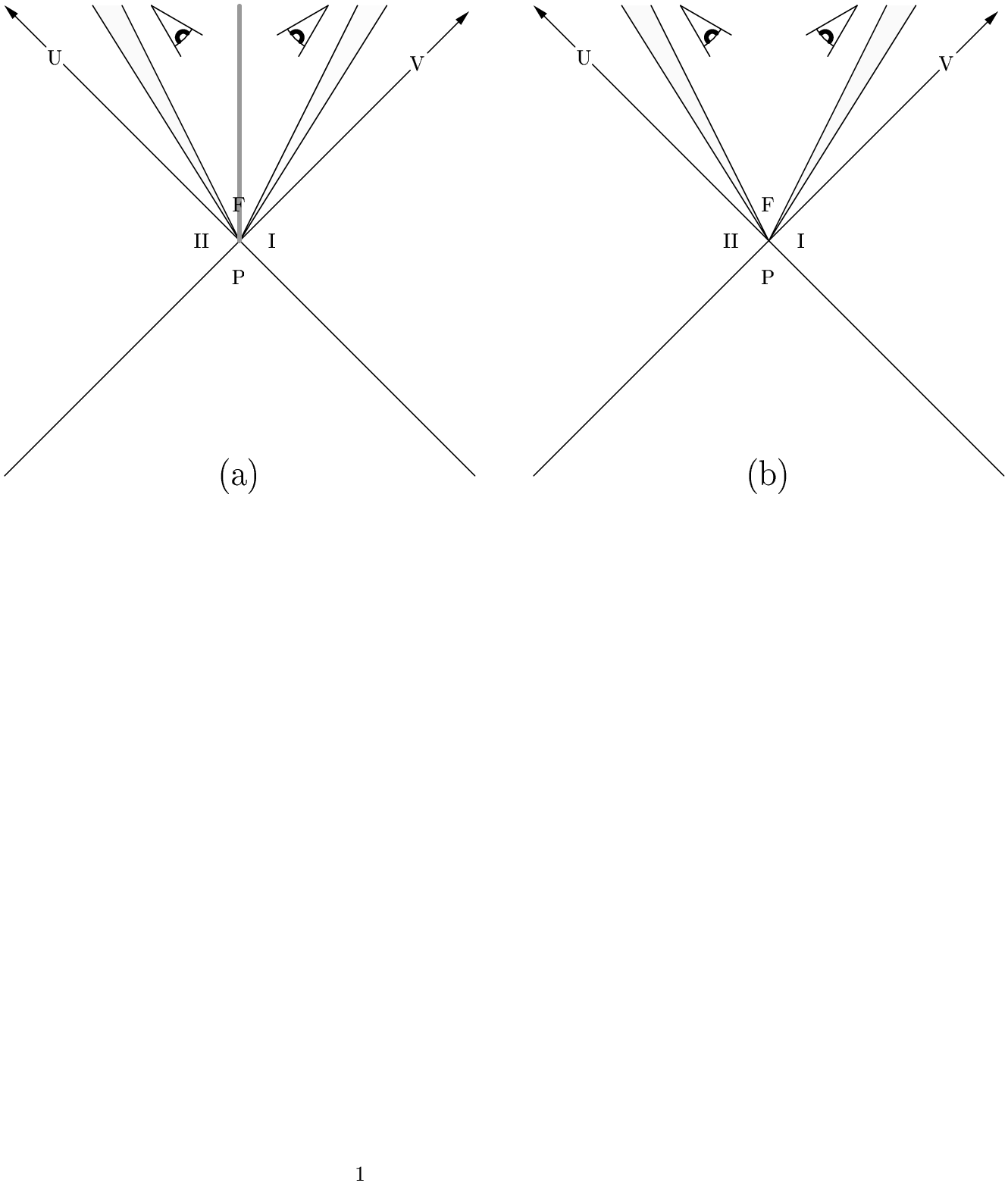}
       \epsfverbosetrue
\caption{Pair of monochromatic detectors (located at $\tau=c_1$ and 
$\tau=c_2$) configured to measure interference
between waves from $I$ and $II$, Figure (a), or to measure the
amplitude of each waves from $I$ and $II$ separately, Figure (b)
(``delayed choice experiment'').
The detectors are monochromatic because they are surrounded by 
interference filters. Each such
filter consists of a Fabry-Perot cavity with uniformly
moving walls. Such a cavity has discrete transmission resonances at 
Rindler frequencies given by $\omega=n\pi /\tanh^{-1}\beta,~~n=1,2,\cdots$,
where $\beta$ is the relative velocity of the cavity walls.
Their histories delineate the evolution of a cavity's interior
(lightly shaded wedge in the figure).}
\label{fig:detectors}
\end{figure}

\noindent (b) places between them a free-float mirror
with freely chosen reflective properties. See Figure~\ref{fig:detectors}.

A brief parenthetical remark is in order: When we say ``inertial
physicist in $F$'', ``free-float particle detector in $F$'', and so
on, what we mean is that the separation between their straight world
lines increases at a constant rate. Moreover, at each instant of their
common proper elapsed time ($0<\xi$) these entities are situated on a
spacelike hypersurface whose points are $(\tau,y,x)$
($-\infty<\tau,y,x<\infty$), and whose neighborhoods are metrically
equivalent (``isometric'') to one another.

The chosen reflectivity of the mirror determines what the detectors
will measure. If the reflectivity is between zero and one
(Figure~\ref{fig:detectors}a), then the detectors respond to the
interfering amplitudes of the waves coming from $I$ and $II$.  If the
reflectivity is zero (mirror is perfectly transparent,
Figure~\ref{fig:detectors}b), then the detectors respond to the
respective amplitudes of the waves coming separately from $I$ and $II$.

The ``delayed choice'' feature enters when the wave intensity is so
low that the individuality of the quanta occupying the wave starts to
manifest itself. Under this circumstance a quantum could be found
still propagating in $I$ and/or $II$, while the physicist is making
up his/her mind whether or not to place a mirror between the two
particle detectors in Figure~\ref{fig:detectors}.

Like its Euclidean relative, the Lorentzian Mach-Zehnder
interferometer $(I,II,P,F)$ is an amplitude splitting device. It also
has the advantage that the two paths for the two amplitudes traverse
regions which are spacious and far apart. This is why it is of
interest to the physicists. These regions, Rindler $I$ and Rindler
$II$, can, for example, accommodate a pair of accelerated dielectrics
or a gravitational disturbance. Either one would alter the
interference between the amplitudes when they recombine in Rindler
sector $F$. As a result antiparticles are produced. This is
demonstrated in Section X.

The mathematicians find this Lorentzian interferometer appealing for
its mathematical simplicity. They notice that it is a device which is
Lorentz boost-invariant in the absence of gravitational disturbances.
As a consequence, the evolution process of an arbitrary wave function,
making its way from $P$ to $F$ via $I$ and $II$, is by necessity
simply a superposition of elementary processes. Each one of them is
easy to follow because each one is a boost-invariant Cauchy evolution
of the Klein-Gordon equation
\begin{equation}
{\partial^2 \psi \over \partial t^2}-
{\partial^2 \psi \over \partial z^2}+ k^2\psi =0,
\label{eq:1}
\end{equation}
where $k^2=k^2_x +k^2_y +m^2c^2/\hbar^2 $. Physically speaking, $\psi$
is the electric (resp. magnetic) field perpendicular to the $x$ and
$y$ directions of the T.M. (resp. T.E.) modes \cite{Candelas and
  Deutsch,Alexander and Gerlach,Higuchi and Matsas} whenever $m=0$.

\subsection{Two Fundamental Sets of Boost-Invariant Modes}

These elementary Cauchy evolutions are fundamental. Each one describes
an independent Mach-Zehnder beam-splitting and (subsequent)
interference process. Furthermore, they serve as a
foundation for carrying the imprints of gravitational disturbances
that might be present in $I$ and/or $II$ \cite{1}.

These Cauchy evolutions are Fourier transforms of the planewave states
\begin{equation}
e^{\mp i [(t-t_0)k\cosh \theta +(z-z_0)k\sinh \theta]}
\equiv e^{\mp ik(Ve^\theta +Ue^{-\theta})/2}
\label{eq:planewave}
\end{equation}
of either positive (upper sign) or negative (lower sign) Minkowski frequency.
These states, we recall, are parametrized by the ``pseudo'' angular parameter 
$\theta$ (``Minkowski frequency parameter'') on the mass
hyperboloid $\omega ^2_k-k^2_z=k^2_x+k^2_y + m^2c^2/\hbar^2 \equiv k^2$,
\begin{eqnarray}
\omega _k&=k \cosh \theta >0 \label{eq:Minkowski frequency}\\
     k_z &=k \sinh \theta \quad ,
\label{eq:z-wave number}
\end{eqnarray}
and we are introducing for the sake of mathematical efficiency
\begin{eqnarray}
U=&(t-t_0)-(z-z_0) \label{eq:retarded}\\
V=&(t-t_0)+(z-z_0)\quad ,
\label{eq:advanced}
\end{eqnarray}
the familiar null coordinates.

These elementary Cauchy evolutions, call them $B^\pm_\omega $, are partial
waves. They are defined so that their superposition
\be
\int^\infty_{-\infty} B^\pm _\omega (kU,kV) e^{-i\omega \theta}~d\omega
\equiv e^{\mp i [(t-t_0)k\cosh \theta +(z-z_0)k\sinh \theta]}
\label{eq:wave expansion}
\ee constitutes a partial wave representation of each planewave mode,
Eq.(\ref{eq:planewave}). Consequently, these partial wave modes are
Fourier transforms on the mass hyperboloid and they come in pairs,
\begin{equation}
B^\pm_\omega (kU,kV)={1\over {2\pi}} \int\limits^{\infty}_{-\infty}
       e^{\mp ik(Ve^\theta +Ue^{-\theta})/2}  e^{i\omega\theta} d\theta 
\quad ,
\label{eq:M-B mode}
\end{equation}
one of purely positive (upper sign), the other of purely negative (lower sign)
Minkowski frequency.
\begin{figure}
\epsfclipon
\epsfysize=7in
\centerline{\epsffile[0 100 612 670]{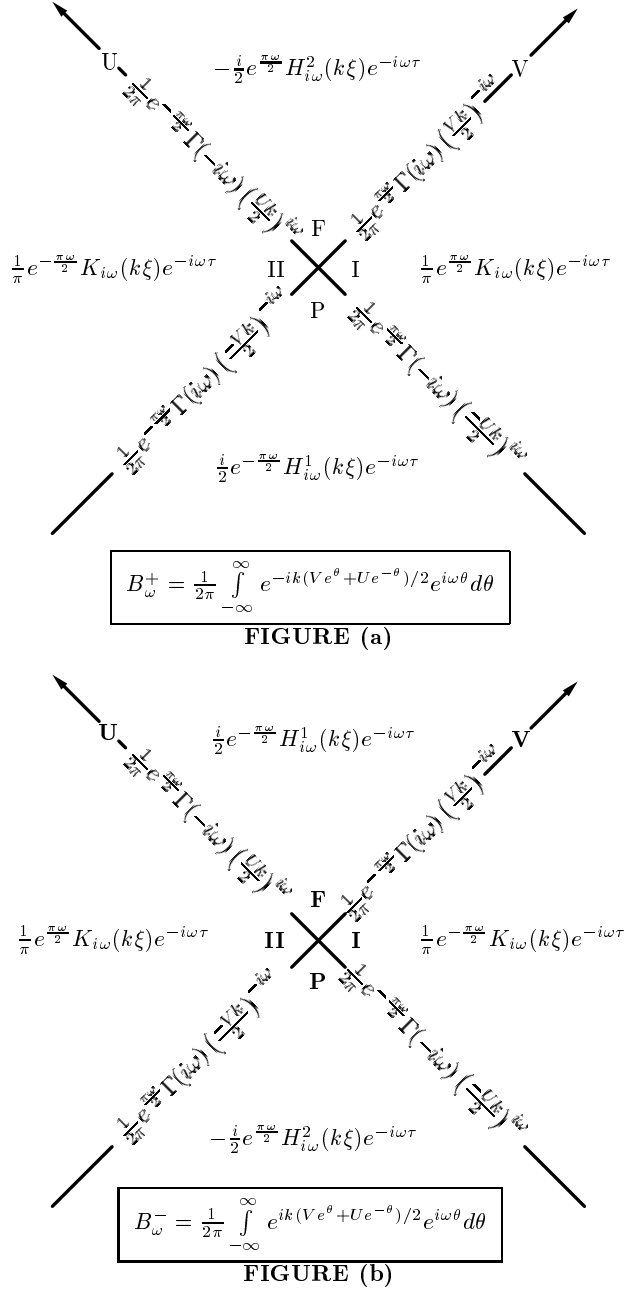}}
       \epsfverbosetrue
\caption{ Minkowski-Bessel modes of positive and negative Minkowski frequency,
and the four Rindler coordinate representatives for each.
Note that the ensuing Figures~\ref{fig:amplitude splitting},
\ref{fig:reflected modes}, and \ref{fig:WKB modes} are mathematically equivalent to the above
figures (Figs.~\ref{fig:M-B modes}a and b). All of them depict
the Minkowski-Bessel modes $B^\pm_\omega$. Figure \ref{fig:reflected modes}
is obtained by applying to Figure~\ref{fig:M-B modes} some 
well-known identities between Hankel and Bessel functions and between
McDonald and modified Bessel functions. Figures \ref{fig:amplitude splitting} 
and \ref{fig:reflected modes} are equivalent by virtue
of the coordinate transformations in Figure \ref{fig:Rindler spacetime}.
Figure \ref{fig:WKB modes} is the WKB-approximate form of Figure \ref{fig:amplitude 
splitting}.}
\label{fig:M-B modes}
\end{figure}
Both are boost invariant, that is to say, 
they are eigenfunctions of the ``boost energy'' operator
\begin{equation}
i{\partial \over {\partial \tau}}\equiv 
i\left[ (z-z_0){\partial \over {\partial t}}+(t-t_0){\partial \over {\partial z}} \right] =
i\left[ V{\partial \over {\partial V}}-U{\partial \over {\partial U}} \right]
\quad 
\end{equation}
with eigenvalue $\omega$. Furthermore, as one sees from 
Figures~\ref{fig:M-B modes}a~and~\ref{fig:M-B modes}b, in each of the 
Rindler sectors they are represented by the various
kinds of Bessel functions.  For this reason we shall refer to them
as ``Minkowski-Bessel'' states or modes or distributions,
depending on the context.

Given their fundamental role, we must make sure that the two kinds of
Minkowski-Bessel states are well-defined physically {\it and}
mathematically. 

\begin{figure}
\epsfclipon
\centerline{\epsffile[0 320 612 680]{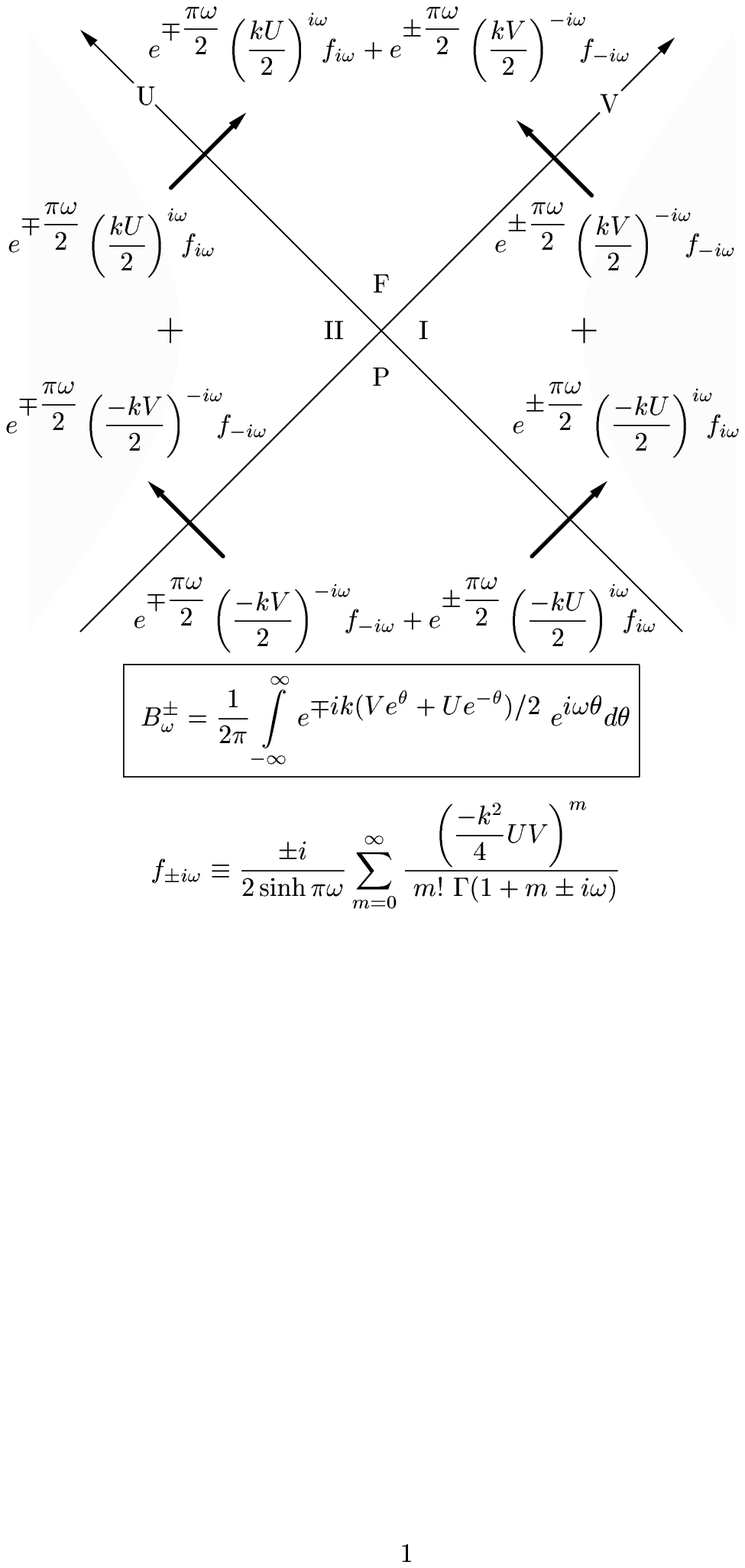}}
       \epsfverbosetrue
\caption{ Amplitude splitting of the Minkowski-Bessel modes of 
positive and negative Minkowski frequency. Each mode expresses an interference
process as found in the Mach-Zehnder interferometer familiar from optics.
A wave in $P$ \emph{far} from the
bifurcation event enters the ``interferometer'' from $P$.
As indicated by the two heavy arrows, the wave splits
into two partial waves: one propagates from $P$ (tail of the arrow)
across the past event horizon, and enters Rindler Sector $I$ (tip of
the same arrow). There it propagates to the right
under the influence of the familiar boost-invariant pseudo-gravitational 
potential. This is shown explicitly in Figure~\ref{fig:reflected modes}. 
The wave becomes evanescent (exponentially decaying)
in the shaded hyperbolic region and hence gets reflected
by this potential. Upon propagating to the left the wave
escapes from Rindler Sector $I$ (tail of the next arrow)
through its future event horizon into the future sector $F$ 
(tip of that arrow).
There it recombines with the other partial wave which got reflected 
in Rindler Sector $II$.}
\label{fig:amplitude splitting}
\end{figure}

Note that the modes $B^\pm_\omega$ are defined globally on Minkowski
spacetime, but as shown in Figures \ref{fig:M-B modes}a and 
\ref{fig:M-B modes}b they have Rindler
coordinate representatives which are only defined in the interior of
each Rindler sector, where they satisfy
\begin{equation}
\left.
\left( \xi \frac{\partial}{\partial \xi}\xi \frac{\partial}{\partial \xi}
-\frac{\partial ^2}{\partial \tau} \mp k^2 \xi^2\right)\psi=0
~~~\right\{
  \begin{array}{l}
  -:~~\hbox{``Rindler Sector I or II''} \\
  +:~~\hbox{``Rindler Sector F or P''} \quad ,
  \end{array}
\label{eq:Rindler equation}
\end{equation}
where, as a consequence, one finds in Rindler $I$ and $II$ the
``pseuso-gravitational potential'' ($\propto k\xi$) already mentioned 
several times, and where
these coordinate representatives are proportional to one or the other
of the two kinds of Hankel functions ($H^{(1)}_{i\omega}$ or
$H^{(2)}_{i\omega}$ in $P$ and $F$) or the MacDonald 
function ($K_{i\omega}$ in $I$ and $II$). On the
acceleration induced event horizons, the boundary ($UV=0$) of these
sectors, the Rindler time coordinate $\tau$ becomes pathological and
so do these functions: they oscillate so rapidly that they have no
definite value on the event horizons $\xi=0$. This is true even when
one uses the well behaved null coordinates $U$ and $V$. See
Figure~\ref{fig:amplitude splitting}.

For obvious reasons such a ``pathological'' behavior seems to be
troubling, both physically and mathematically. In fact, the literature,
as well as discussions with workers in the field, have emphasized this
pejorative assessment by implicitly and explicitly describing the
behaviour of the these Minkowski-Bessel modes across the event horizon
as being ``singular'' or ``highly singular''. All this has lent 
credence to the misconception that these modes are physically
unrealistic and mathematically unjustified ways of coming to grips with 
natural phenomena associated with pairs of accelerated frames.

Furthermore, the highly oscillatory (non-analytic) behaviour of
$B^\pm_\omega (kU,kV)$ at $U=V=0$ has been used, erroneously, as an
indicator to signal the mixing of positive and negative Minkowski
frequencies\cite{Davies}.

The purpose of this article is to dispel these mathematical and
physical misconceptions. This turns out to be a fruitful endeavor for
additional reasons, which are theoretical and (hence also) practical:
By placing the Minkowski-Bessel modes into their wider mathematical
framework one can assess their status in our hierarchy of
knowledge. Furthermore, this wider framework also leads to a complete
set of globally well-behaved orthonormal wavepackets, which (i) serve as
the connecting link between quantum mechanical wave functions and the
razor-sharp world lines of classical mechanics and (ii) open the door to
probing the structure of spacetime with probes (exploding wave packets,
Section VII.C.2) which are as yet unexplored.

\section{ TWO COMPLEMENTARY POINTS OF VIEW}

The Minkowski-Bessel modes oscillate very rapidly near the event
horizon. The erroneous assessment of this behavior as ``pathological''
is tied up in their very description as ``modes'' and is a direct
consequence of the concomitant and frequently asked question: ``What
is the behavior of one of these modes at the origin $U=V=0$ where the
two event horizons intersect?''  or ``What is its behavior as one
passes across one of the event horizons $\vert t-t_0\vert = \vert
z-z_0\vert$?''  These are questions that pertain to the continuity or
smoothness of the normal mode amplitudes.

\begin{figure}[!tbp]
\epsfclipon
\epsffile[200 370 712 600]{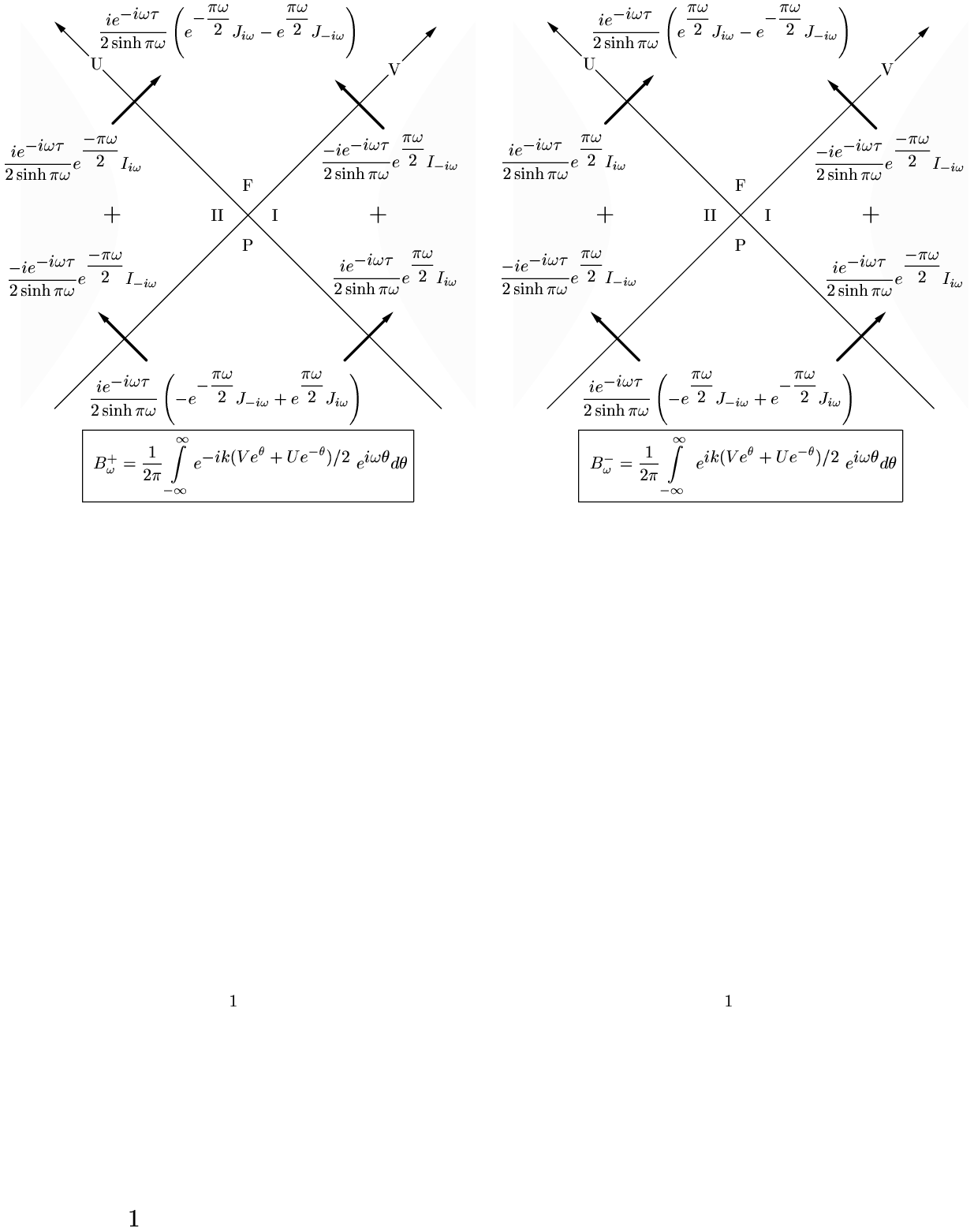}
       \epsfverbosetrue
\caption{Reflection of partial waves from their potential barriers
(shaded hyperbolas) in Rindler Sectors $I$ and $II$. As in
Figure~\ref{fig:amplitude splitting}, the four arrows connect partial waves
propagating from one Rindler sector to another.  In Rindler Sectors
$P$ and $F$ the partial waves are proportional to the Bessel functions
$J_{\pm i \omega}(k\xi)$ of order $\pm i \omega$, while in $I$ and $II$
they are proportional to the modified Bessel functions $I_{\pm
i\omega}(k\xi)$. The latter oscillate for finite $k\xi$, but blow up
exponentially as $k\xi \rightarrow \infty$ in each hyperbolically
shaded region. The reflection process is brought about by the boundary
condition that the incident wave in $I$ ($\propto I_{i \omega}(k\xi)$)
must combine with the reflected wave ($\propto I_{-i \omega}(k\xi)$) so
as to form a total wave whose amplitude approaches zero as $k\xi
\rightarrow \infty$. This wave is a standing wave ($\propto (I_{i
\omega}(k\xi) - I_{-i \omega }(k\xi))$ in $I$ and thus expresses a
process of reflection from the boost-invariant pseudo-gravitational 
potential.
An analogous reflection process takes place in Rindler Sector $II$.}
\label{fig:reflected modes}
\end{figure}

\begin{figure}[h!]
\epsfclipon
\epsffile[60 400 712 650]{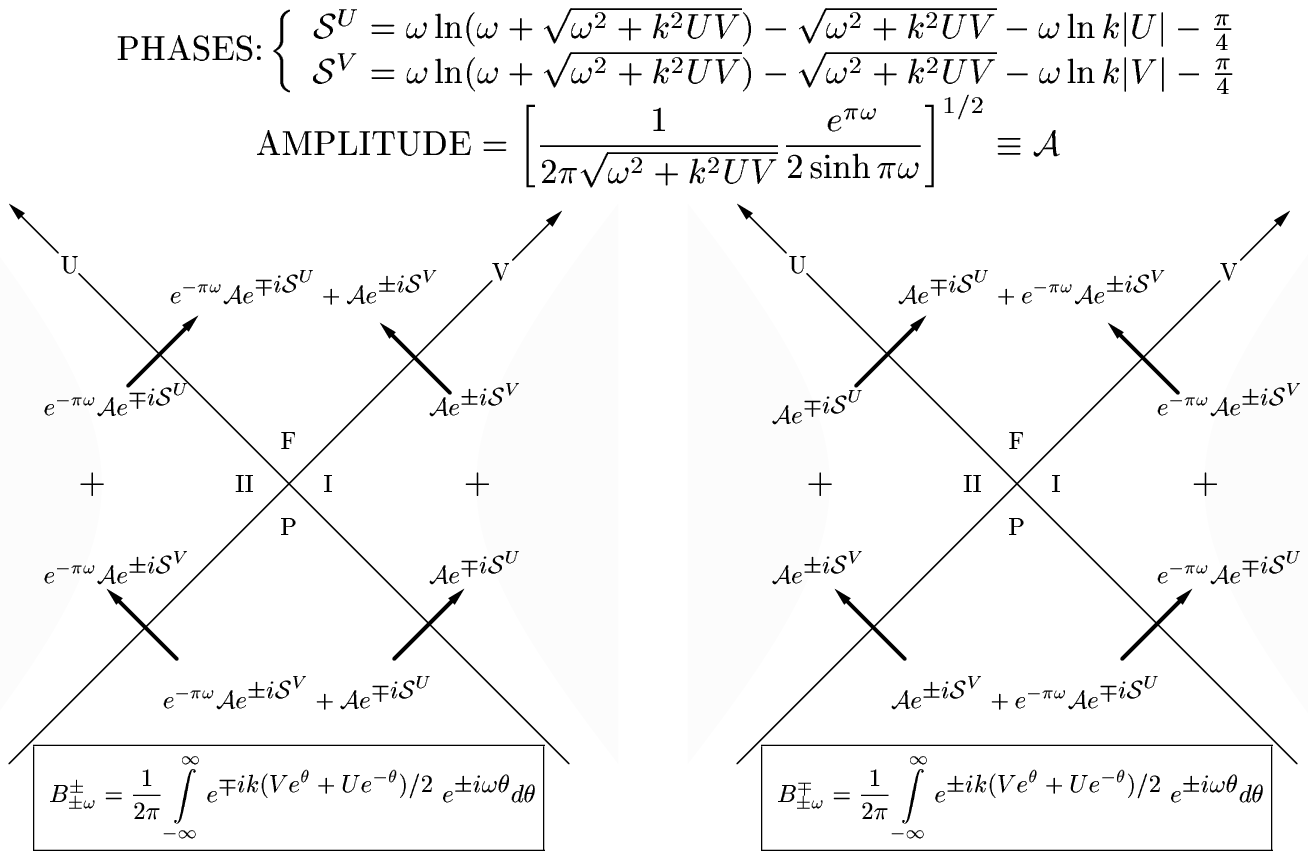}
       \epsfverbosetrue
\caption{WKB-approximate formulas for the Minkowski-Bessel modes. All
formulas assume that $\omega=\vert \omega \vert$. The upper (lower)
signs in a picture go only with the upper (lower) signs in the corresponding boxed
formula. This means, for example, that the formula (in all four
Rindler sectors) for a M-B of, say, positive Minkowski frequency and of
negative Rindler frequency, is designated by $B^+_{-\omega}$, and is
given by the lower signed expressions in the second picture.
The complete propagation process (splitting, reflection, and
interference) of a M-B mode is characterized by the phase ${\mathcal S}^U$, 
which
is continuous across the $U$-axis, and the phase ${\mathcal{S}}^V$, which is
continuous across the $V$-axis. Both satisfy the Hamilton-Jacobi
equation 
$\frac{\partial S}{\partial U}\frac{\partial S}{\partial V}=-\frac{k^2}{4}$, 
which is implied by Eq.(\ref{eq:1}). The amplitude
$\mathcal{A}$ or ($e^{-\pi\omega} \mathcal{A}$) of each WKB wave propagating across its respective event horizon satisfies the concomitant conservation law
$\frac{\partial}{\partial U}\left( {\mathcal{A}}^2\frac{\partial S}
{\partial V}\right)
+\frac{\partial}{\partial V}\left( {\mathcal{A}}^2\frac{\partial S}
{\partial U}\right)=0$, 
where $S={\mathcal{S}}^U$ or ${\mathcal{S}}^V$, depending on which horizon 
is being crossed. 
}
\label{fig:WKB modes}
\end{figure}

Questions like these are a consequence of the {\it conventional
viewpoint}. It regards $B^\pm_\omega (kU,kV)$ given by
Eqs.(\ref{eq:M-B mode}) and depicted in Figures \ref{fig:M-B modes}a 
and \ref{fig:M-B modes}b (or in
Figures~\ref{fig:amplitude splitting}, \ref{fig:reflected modes}, and 
\ref{fig:WKB modes}) 
as ``normal modes'', and thereby asks
for their amplitudes as one varies $(U,V)$ while keeping the parameter
$\omega$ fixed.

This viewpoint will suffice if one's attention is limited to the
interior of anyone of the four Rindler sectors $I,II,F,$ or $P$. In
that case one has four ordinary functions, the coordinate
representatives of $B^+_\omega (kU,kV),$ and another four for
$B^-_\omega (kU,kV)$, as depicted in 
Figures~\ref{fig:M-B modes}a and \ref{fig:M-B modes}b.

However, we also need to know the amplitude on the boundary
(event horizons) between the Rindler sectors. For that, the conventional
viewpoint is deficient in that it forces us not only into asking
mathematically meaningless questions, but also into referring to
``properties'' of the field which are physically immeasurable.

By contrast, the {\it complementary viewpoint} regards
$B^\pm_\omega (kU,kV)$ as functions of $\omega$ and views $U$ and $V$
as parameters.  The superiority of this viewpoint lies
in that it allows us to extend our familiar notion of ordinary
functions to {\it generalized} functions, which are well-defined even
when $(U,V)$ lies on the boundary.

These generalized functions (``distributions'') (i) subsume ordinary
functions (of $\omega$) as a special case, and (ii) are continuous and
smoothly parametrized not only when $(U,V)$ lies on the boundary
between any pair of Rindler sectors, but also when $U=V=0$, the
intersection of the two event horizons. Furthermore, unlike a
(monochromatic) normal mode, a generalized function expresses
mathematically what is measured physically.

\section{ARE GENERALIZED FUNCTIONS PHYSICALLY NECESSARY?}

Before identifying the two families of generalized functions we must
ask: What fact(s) of reality give rise to these generalized functions?
i.e., Do they have a natural origin, and if so, what is it? A brief
answer lies in the following two observation:

First of all, in nature one does not find translation and boost
eigenfunctions (of infinite extent and infinitely sharp momentum) as
such.  Instead, one only observes wave packets of finite (although
possibly quite large) extent and of fuzzy (although possibly quite
sharp) momentum. See Section VII.  Nevertheless, translation and boost
invariance {\it are} physical properties found in nature, and for good
reasons it is necessary to express them in terms of translation and
boost eigenfunctions.  In order to harmonize the conflict between
this necessity and what is observed, physicists and mathematicians have
enlarged the concept of an eigenfunction into the more sophisticated
concept of a {\it generalized eigenfunction}, i.e. an {\it
eigendistribution}.  Thus for physical and mathematical reasons the
translation and boost invariance in nature must actually be formulated
in terms of generalized functions.  Ordinary functions won't do.
Without generalized functions there would not exist, for example, 
a globally defined
solution to the Klein-Gordon equation which is Lorentz invariant.

Secondly, there is an equally, if not more, urgent reason for introducing them:
The general solution to the Klein-Gordon equation is a linear combination
of the two sets of Minkowski-Bessel modes, Eq.(\ref{eq:M-B mode})
\be
\psi =
\int\limits^{\infty}_{-\infty} [\phi ^+(\omega)~ B^+_\omega (kU,kV)
+ \phi ^-(\omega)~ B^-_\omega (kU,kV)]  ~ d \omega \quad .
\label{eq:solution}
\ee
The two functions $\phi ^\pm(\omega)$ comprise an $SU(1,1)$ two-spinor
field\cite{1} over the Rindler frequency domain $-\infty < \omega < \infty $.
It is difficult to overstate the importance of this spinor field. Its Stokes
parameters (= expectation values of the Pauli matrices)
form a vector field which (a) carries the imprints of gravitation, and (b)
has the Eotvos property of being independent of the mass of the Klein-Gordon
particle\cite{1}. 
This fundamental role of the spinor field demands that the 
solution represented by Eq.(\ref{eq:solution}) be globally well-defined, 
including at events on the event horizon $UV=0$. 

Thus the motivation for considering the two Minkowski-Bessel
distributions and their test function spaces comes not only from
mathematical analysis but also from physics and geometry.

\section{MATHEMATICAL TUTORIAL: TWO FAMILIES OF GENERALIZED FUNCTIONS}

Generalized functions are a familiar part of the mathematical 
landscape\cite{Gelfand and Shilov}.
Nevertheless, there seems to be lacking a readily accessible 
mathematical framework
which accommodates the boost-invariant and globally defined eigenmodes
of the Klein-Gordon equation. The purpose of this section is to remedy
this lack.

\subsection{Space of Test Functions}

Recall that, unlike ordinary functions, generalized functions are not
defined by themselves, but depend on the space of test functions.
This means that a generalized function is a linear functional on the
space of test functions. The space of test functions can be one of
several possibilities, but the one we are mainly interested in consists of
those functions {\it which are the Fourier transforms of functions compact
support}. The formula for such test functions is
\begin{eqnarray*}
\hat \phi (\omega)
&=&\frac{1}{\sqrt{2\pi}}\int\limits^{\infty}_{-\infty}\phi (\theta) 
                        e^{i\omega\theta} d\theta \\
&=&\frac{1}{\sqrt{2\pi}}\int\limits^{a}_{-a}\phi (\theta) 
                        e^{i\omega\theta} d\theta
\end{eqnarray*}
because the support of $\phi (\theta)$ is confined to the compact region 
$\vert \theta \vert \le a$. 
These test functions, $\hat \phi(\omega)$,
are entire analytic
functions of their argument $z=\omega +i \zeta$, and
they satisfy the inequality
\be
\vert \hat \phi (\omega +i \zeta) \vert \le C e^{a\vert \zeta \vert } \quad .
\label{eq:space of entire functions}
\ee

Adapting the notation of Gelfand and Shilov
\cite{Gelfand and Shilov}, 
we shall call the space
of such test functions $Z_0$. This is the space of test functions whose
inverse Fourier transforms, the space $K_0$, are of compact support. 
Thus $\phi$ belongs to $K_0$, while the Fourier transform of $\phi$, the
test function $\hat \phi$, belongs to $Z_0$.  
The subscript zero
designates the fact that each $\phi$ is piecewise continuous,
while $Z_1$ would be the space of Fourier transforms of continuous
functions with compact support, $Z_2$ the space of Fourier transforms
of functions with continuous first derivatives and with compact support, 
etc., and $Z_\infty \equiv Z$ the Fourier transform of $K_\infty \equiv K$,
the space of infinitely differentiable (``smooth'') functions 
with compact support.

\subsection{Spacetime-Parametrized Family of Generalized Functions}

Generalized functions fall into two mutually exclusive and jointly exhaustive
types: {\it regular} and {\it singular}. We shall need to consider both.

A {\it regular} generalized function is a linear functional on $Z_0$
defined by the integral of an ordinary (and locally integrable) function
multiplied by a test function in $Z_0$.

Accelerated frames give rise to two regular $(U,V)$-parametrized
families of generalized functions. They are defined by the linear
functional (asterisk $^*$ denotes complex conjugate)
\be
 \int\limits^{\infty}_{-\infty}  \hat \phi (\omega)^*~  B^\pm_\omega (kU,kV)~ d \omega
\equiv (\hat \phi,B^\pm(kU,kV)) 
\quad 
\label{eq:linear functional}
\ee
on the test functions $\hat \phi(\omega)$, which belong to $Z_0$. The function
$B^\pm_\omega (kU,kV)$ is represented by the integral
Eq.(\ref{eq:M-B mode}), which is an ordinary function of $\omega$ 
whenever $(U,V)$
lies in the interior of the four Rindler sectors $I,II,P,$ or $F$.

{\it Notational Remark.} Even though there are {\it two} families,
{\it two} integrals, {\it two} generalized functions, and so on, we
shall from now on refer to the two collectively in the singular, and
not in the plural, and let the plus and minus sign serve as labels
that refer to one or the other.

The integral representatives for $B^\pm_\omega (kU,kV)$
are exhibited in Figures~\ref{fig:M-B modes}a and \ref{fig:M-B modes}b 
for each of the four Rindler
sectors.  The requisite relation between the Rindler coordinates
$(\tau,\xi)$ and the null coordinates $(U,V)$ in these four Rindler
Sectors is shown in Figure 1.

The space of all linear functionals on $Z_0$ forms a linear space in its
own right, and its name is $Z'_0$. The set of regular generalized functions
\be
\{B^\pm (kU,kV):~(U,V)~\hbox {lies~inside~one ~of~the~Rindler~Sectors}\}
\ee
is a subset of $Z'_0$. More precisely, let
\be
R=I\cup II\cup P \cup F
\label{eq:interior}
\ee
be the union of the interior of the four Rindler quadrants. Then the
two-parameter family $B^\pm(kU,kV)$ is a continuous map from $R$ into
$Z'_0$, the space of all generalized functions associated to $Z_0$:
\be
B^\pm ~:~R~~\longrightarrow ~~Z'_0 \quad. 
\label{eq:distribution}
\ee
In fact, the map is a smooth map because all its derivatives with respect to
$U$ and $V$ exist in its whole domain, $I\cup II\cup P \cup F$

However, when $(U,V)$ approaches the event horizon, a point on the
boundary of $R$, then the $(U,V)$-parametrized {\it ordinary} function
$B^\pm_\omega (kU,kV)$ does not tend towards an ordinary function. In
fact, the limit does not even exist. Nevertheless, we shall see that
the integral, Eq.(\ref{eq:linear functional}), does tend towards a 
limit for all test
functions in $Z_0$. Thus the generalized function $B^\pm(kU,kV)$ has a limit
as $(U,V)$ approaches the event horizon, even though the ordinary
function $B^\pm_\omega(kU,kV)$ does not.

\subsection{Extension Onto the Event Horizon}

Each of the null coordinates $U$ and $V$ is smooth across the event
horizon. However, the $(U,V)$-parametrized family of generalized
functions $B^\pm (kU,kV)$ is not, because it is not even defined
there.  What we need is an {\it extension} of this family across the
event horizon. This means that this extended two-parameter family,
call it $\overline B^\pm (kU,kV)$, is the same as the original family
$B^\pm (kU,kV)$, except that the parameter domain of $\overline B^\pm
(kU,kV)$ is bigger than that of $B^\pm (kU,kV)$.

\subsubsection{Definition}
The parameter domain of the to-be-constructed extension 
should be {\it all} of 
Minkowski spacetime,
\be
\overline R
=~I\cup II\cup P \cup F \cup \{(U,V):UV=0\} \quad ,
\label{eq:extended region}
\ee
which includes also the event horizons
\be
\{(U,V):~UV=0\} \quad ,
\label{eq:event horizon}
\ee
the boundary of the four Rindler Sectors $I,II,P,$ and $F$.
Thus the to-be-constructed extension is a map from $\overline R$ into $Z'_0$,
\be
\overline B^\pm ~:~\overline R~~\longrightarrow ~~Z'_0 \quad. 
\label{eq:extension}
\ee
with the property that
\be
\overline {B}^\pm(kU,kV)=B^\pm(kU,kV)~~ \hbox{ whenever} ~~(U,V)\in R
\quad; 
\ee
that is $\overline {B}^\pm\vert _R =B^\pm$.  

A map, such as Eq.(\ref{eq:extension}), is said to be an {\it extension} of 
the map
Eq.(\ref{eq:distribution}), because $R\subset \overline R$ and $\overline 
{B}^\pm\vert _R =B^\pm$.

An extension of the two-parameter family $B^\pm(kU,kV)$,
Eq.(\ref{eq:linear functional}), to the event horizons,
Eq.(\ref{eq:event horizon}), is not arbitrary. In fact, it is (i)
unique, (ii) continuous, and (iii) even smooth. This, we shall see, is
because this extension is related by the Fourier transform to the
Minkowski plane wave amplitude on the mass hyperbola,
Eqs.(\ref{eq:Minkowski frequency}), (\ref{eq:z-wave number}).  This
amplitude, Eq.(\ref{eq:planewave}), is a $(U,V)$-parametrized family
of ordinary functions. Its parameter domain includes all of spacetime,
including the event horizons, Eq.(\ref{eq:event horizon}). This fact
is the key to $\overline B^\pm(kU,kV)$, the continuous and smooth
extension of $B^\pm(kU,kV)$.

\subsubsection{Construction}

The construction of $\overline B^\pm(kU,kV)$ is based on a
familiar idea, namely, Parceval's theorem. Let us fix notation
by letting
\begin{eqnarray}
\hat \phi (\omega)
&=& \frac{1}{ \sqrt{2\pi} }\int\limits^{\infty}_{-\infty}\phi (\theta) 
                        e^{i\omega\theta} d\theta \label{eq:Fourier transform} \\
\hat f (\omega)
&=& \frac{1}{\sqrt{2\pi}}\int\limits^{\infty}_{-\infty}f (\theta) 
                                        e^{i\omega\theta} d\theta \nonumber
\end{eqnarray}
be the Fourier transforms of $\phi$ and $f$.
Then,  with the asterisk, $^*$, denoting complex conjugate, one has
\begin{eqnarray*}
(\phi,f) &\equiv & \int\limits^{\infty}_{-\infty}  \phi(\theta)^*f(\theta)~
                                                        d\theta 
        = \int\limits^{\infty}_{-\infty}  \phi(\theta)^*
\left\{ \frac{1}{ \sqrt{2\pi} }
\int\limits^{\infty}_{-\infty} \hat f (\omega)e^{-i\omega \theta} d\omega \right\}
                                                d\theta \\
        &=& \int\limits^{\infty}_{-\infty}  \hat f (\omega)
\left\{ \frac{1}{ \sqrt{2\pi} }
\int\limits^{\infty}_{-\infty} \phi (\omega)e^{i\omega \theta} d\theta 
                                                                \right\}^* d\omega 
        =  \int\limits^{\infty}_{-\infty}  \hat \phi(\omega)^*\hat f (\omega)~
                                                                d\omega \\
&=& (\hat \phi,\hat f) \quad .
\end{eqnarray*}
This is Parceval's theorem.
We now turn this theorem around and use its conclusion,
\be
(\hat \phi,\hat f)=(\phi,f)~,
\ee
to define the generalized function $\hat f$. Its domain is the space 
$Z_0$ of test functions $\hat \phi(\omega)$.
For a given function $f$, the value of this linear functional $\hat f$ is
uniquely determined by the integral on the right hand  side. This is because
$\phi(\theta)$ belongs to $K_0$, the space of functions with compact support
$-a\le\theta \le a$.

Apply this ``inverted Parceval's theorem'' to the function
\[
f(\theta)=\frac{1}{ \sqrt{2\pi} }e^{\mp ik(Ve^\theta +Ue^{-\theta})/2}~,
\]
the plane wave amplitude, Eq.(\ref{eq:planewave}). Denote the resulting linear functional by 
\[
\hat f={\overline B}^\pm(kU,kV)~.
\]
One, therefore, has the following definition 
\be
 (\hat \phi ,{\overline B}^\pm(kU,kV)) \equiv
\int\limits^{\infty}_{-\infty}
    \phi (\theta)^* \left( \frac{1}{ \sqrt{2\pi} }
e^{\mp ik(Ve^\theta +Ue^{-\theta})/2}\right)     d\theta 
\quad .
\label{eq:integral}
\ee
This integral (as well as all its partial derivatives with respect to
$U$ and $V$) is obviously finite because $\phi (\theta)$ has
compact support. In fact, this integral, as well as all its derivatives,
are finite for all parameter values $-\infty < U,V < \infty $, even
for those for which $UV=0$. Furthermore, for every $\hat \phi (\omega)$ in
$Z_0$, this integral depends continuously and smoothly on the
parameter pair $(U,V)$. Consequently, the generalized function
$\overline B^\pm(kU,kV)$ defined by Eqs.(\ref{eq:Fourier
transform},\ref{eq:integral}) is a continuous and smooth function of
the parameters $U$ and $V$, including those $(U,V)$ which lie on the
event horizon.

The usefulness of $\overline B^\pm(kU,kV)$ as defined by
Eqs.(\ref{eq:Fourier transform},\ref{eq:integral}) derives from the
fact that it coincides with $B^\pm(kU,kV)$, Eq.(\ref{eq:M-B mode}),
whenever $(U,V)$ lies in $R$, Eq.(\ref{eq:interior}). If true, this
makes $\overline B^\pm$ the extension of $B^\pm$. This is verified
below. In the next section (Section V) we shall see how this
coincidence leads to a very intuitive integral representation of
$\overline B^\pm(kU,kV)$.

Consider $\overline {B}^\pm(kU,kV)$ as defined by Eqs.(\ref{eq:Fourier
transform},\ref{eq:integral}). The fact that $\phi (\theta)$ is
of compact support, say $-a<\theta <a$, yields
\begin{eqnarray}
        (\hat \phi,\overline {B}^\pm(kU,kV)) 
    &\equiv& \int\limits^{\infty}_{-\infty} \phi (\theta)^*
      \frac{1}{ \sqrt{2\pi} }e^{\mp ik(Ve^\theta +Ue^{-\theta})/2} 
                     d\theta \nonumber\\
    &=&       \int\limits^{a}_{-a} \phi (\theta)^*
        \frac{1}{ \sqrt{2\pi} }e^{\mp ik(Ve^\theta +Ue^{-\theta})/2} 
                     d\theta \nonumber\\
    &=&\int\limits^{\infty}_{-\infty} \hat \phi (\omega)^*\left[
         {1\over {2\pi}} \int\limits^{a}_{-a}
           e^{\mp ik(Ve^\theta +Ue^{-\theta})/2}  e^{i\omega\theta} d\theta 
             \right] d\omega \quad .
\label{eq:double integral}
\end{eqnarray}
This double integral is unchanged if one lets $a\to \infty$. 
Consequently, for $UV\ne 0$ the $\theta$-integral has the limit
\be
\lim_{a\to \infty}~{1\over {2\pi}} \int\limits^{a}_{-a}
       e^{\mp ik(Ve^\theta +Ue^{-\theta})/2}  e^{i\omega\theta} d\theta
\equiv {B}^\pm_\omega (kU,kV) \quad ,
\ee
whose value $B^\pm_\omega (kU,kV)$ inside any one of the Rindler Sectors
is indicated in Figures 4a and 4b.
Thus, Eq.(\ref{eq:double integral}) becomes
\begin{eqnarray}
                 (\hat \phi,\overline {B}^\pm(kU,kV))
              & =&\int\limits^{\infty}_{-\infty}    
                   \hat \phi (\omega)^* B^\pm_\omega (kU,kV)~ d\omega \nonumber \equiv \\
         &\equiv & ( \hat \phi,{B}^\pm(kU,kV)) \quad .
\end{eqnarray}
This holds for all test functions $\hat \phi$. Consequently, we have 
\[ \overline {B}^\pm(kU,kV)=B^\pm(kU,kV)~~ \hbox{ whenever} ~~(U,V)\in R
 \quad; \]
that is $\overline {B}^\pm\vert _R =B^\pm$.  In other words,
$\overline B^\pm(kU,kV)$ {\it is an extension of} $B^\pm(kU,kV)$ indeed.

\subsubsection{Smoothness}
It is clear that the extension $\overline B^\pm(kU,kV)$ is continuous
and smooth in the parameters $U$ and $V$ when $(U,V)$ lies in the
interior of one of the Rindler Sectors; after all, there, by construction, 
$\overline
B^\pm$ coincides with $B^\pm$, which is continuous and smooth in $U$
and $V$.  But $\overline B^\pm(kU,kV)$ is also continuous and smooth
when $(U,V)$ lies on the boundary, Eq.(\ref{eq:event horizon}). 
For example, continuity
of $\overline B^\pm$ is expressed by
\be
\lim_{(U,V)\to(U_0,V_0) } \overline B^\pm(kU,kV) 
=\overline B^\pm(kU_0,kV_0) \quad ,
\ee
while smoothness is expressed by
\be
\lim_{(U,V)\to (U_0,V_0)}~~
{\partial ^{m+n} \over \partial U^m\partial V^n} \overline B^\pm(kU,kV)
= {\partial ^{m+n} \over \partial U^m\partial V^n} \overline B^\pm(kU_0,kV_0)
\label{eq:partial derivative}
\ee
This is obtained by differentiating Eq.(\ref{eq:double integral})
\begin{eqnarray}
\lefteqn{ \left( \hat \phi,
               {\partial ^{m+n} \over \partial U^m\partial V^n} 
                \overline B^\pm(kU,kV)
          \right)  }   \label{eq:smeared partialderivative} \\
        &\equiv&\left(\mp {ik \over 2} \right) ^{m+n}
                \int\limits^{\infty}_{-\infty} \phi (\theta)^*
                \left[\frac{1}{ \sqrt{2\pi} }e^{\mp ik(Ve^\theta +Ue^{-\theta})/2} e^{(n-m)\theta}\right]  d\theta 
                                \label{eq:first explicit partialderivative}\\
        &=&\left(\mp {ik \over 2} \right) ^{m+n}
                \int\limits^{a}_{-a} \phi (\theta)^*
                \left[ \frac{1}{ \sqrt{2\pi} }e^{\mp ik(Ve^\theta +Ue^{-\theta})/2} e^{(n-m)\theta}\right]
                  d\theta    
                                \label{eq:second explicit partialderivative}\\
        &=&\left(\mp {ik \over 2} \right) ^{m+n} \int\limits^{\infty}_{-\infty}
                        \hat \phi (\omega)^*
               \left[
                    {1\over {2\pi}} \int\limits^{a}_{-a}
                     e^{\mp ik(Ve^\theta +Ue^{-\theta})/2} 
                     e^{(n-m)\theta}e^{i\omega \theta}
                     d\theta 
                \right] 
                         d\omega ~~.\nonumber
\end{eqnarray}
This expression depends continuously on the parameter pair $(U,V)$, even
when $(U,V)$ lies on one of the event horizons where $UV=0$. This is
guaranteed by the finiteness of the the integral in Eq.(\ref{eq:second 
explicit partialderivative}). Consequently, one has 
\be
\lim_{(U,V)\to (U_0,V_0)}~~\left(\hat \phi,
{\partial ^{m+n} \over \partial U^m\partial V^n} \overline B^\pm(kU,kV)
\right)
    =\left( \hat \phi,
           {\partial ^{m+n} \over \partial U^m\partial V^n} 
           \overline B^\pm(kU_0,kV_0)
     \right)~~.
\ee
This holds for all test functions $\hat \phi(\omega)$ belonging to $Z_0$,
and thus verifies the smoothness condition
(\ref{eq:partial derivative}) at all events $(U_0,V_0)$ of Minkowski
spacetime.

\subsection{Other Test Function Spaces} 
This smoothness raises two pertinent questions: Are there any other test
function spaces, besides $Z_0$, relative to which $\overline
B^\pm(kU,kV)$ depends continuously on the parameters $U$ and $V$? 
and: What are these test function spaces good for?

We answer the last question first by noting that these spaces supply
us with complete sets of orthonormal basis functions which are
square-integrable on the real line. They are exhibited for the first
two examples below. From these sets one obtains quite
trivially complete sets of Klein-Gordon-orthonormal wavepacket
histories, the dynamical degrees of freedom of the Klein-Gordon
system. Their construction is illustrated in Section VII.

The answer to the first question is ``yes'', there are other test
function spaces. What are they?
That depends entirely on the behavior which $\overline B^\pm(kU,kV)$
is required to have. This means that one talks about \emph{the properties}
of the generalized function $\overline B^\pm(kU,kV)$ \emph{relative to 
the given space of test functions}.

1.) As a first example, consider $Z_0$ as was done above. One says
that $\overline B^\pm(kU,kV)$ is smooth (i.e. infinitely
differentiable) \emph{relative to the given space of test functions
$Z_0$}. The functions of this space are very useful because they and
their Fourier transforms accommodate orthonormal bases which, as shown
in Section VII.B, lead to sets (i.e. bases) of
Klein-Gordon-orthonormal wave packet histories.

However, a cursory examination of these wave packet solutions shows
that they are based on orthonormal test functions whose Fourier
transform is discontinuous on the $\theta$-domain, the mass
hyperboloid Eqs.(\ref{eq:Minkowski frequency}), (\ref{eq:z-wave
number}).  To ward off such a potentially unphysical assumption, we
shall exhibit a set of test functions whose Fourier
transform is smooth on the $\theta$-domain, \emph{and} relative to which the
Minkowski-Bessel distribution yields wave packet histories which are
still smooth everywhere, including the event horizons. We shall do
this in the next example.

2.) As a second example, consider the space of smooth test functions
which decay faster than any exponential. We shall denote this space by
$G$ because it accommodates the set of functions with Gaussian decay,
\be
\phi(\theta)=H_n(\theta)e^{\displaystyle -\theta ^2 /2}~,~~~~n=0,1,2,\cdots~~,
\ee
where $H_n(\theta)$ is the Hermite polynomial of order $n$.
These Gaussian test functions (``simple harmonic oscillator eigenfunctions'')
have the delightful property that they are eigenfunctions of the 
unitary Fourier transform:
\be
\int_{-\infty}^\infty e^{\displaystyle -\theta ^2 /2} H_n(\theta)
\frac{e^{i\omega \theta }}{\sqrt{2\pi} } ~d\theta
=i^n~e^{\displaystyle -\omega ^2/ 2} H_n(\omega)~.
\ee
The eigenfunction property of these Gaussian test functions secures the
best of both worlds: (1) faster than exponential decay on the 
Minkowski frequency $\theta$-parameter domain \emph{and} (2)
smoothness both on the Rindler frequency ($\omega$) domain and the 
Minkowski frequency $\theta$-domain. 

Thus we have achieved our goal of exhibiting a set of test functions
which together with their Fourier transforms are smooth \emph{and}
which guarantee the convergence of the integral,
Eq.(\ref{eq:first explicit partialderivative}), and hence guarantee
the continuity of all spacetime derivatives, Eq.(\ref{eq:partial
derivative}), of the Minkowski-Bessel distribution.

3.) As a third, but less important example, consider the space $K_0$
consisting of test functions $\hat \psi(\omega)$ with compact support
on the Rindler frequency line $-\infty <\omega <\infty$.  This test
function space is very different from $Z_0$ of example~1. In fact,
with $Z_0$ consisting only of entire analytic functions, these two
spaces have no common element except the zero function.  (A non-zero
function of compact support cannot be analytic.)  It is quite easy to
see that, relative to the test function space $K_0$, the extension
$\overline B^\pm(kU,kV)$ is continuous (but not smooth) in the
parameters $U$ and $V$. Indeed, consider the (inverse) Fourier
transform
\be
\psi (\theta)
= \frac{1}{ \sqrt{2\pi} }\int\limits^{\infty}_{-\infty}\hat \psi (\omega) e^{-i\omega\theta} d\omega
\quad
\ee
of $\hat \psi(\omega)$, which is a function of compact support. 
This (inverse) Fourier transform
is a function which decays at least as rapidly as ${1\over
\vert \theta \vert}$ along the $\theta$-axis. Consequently,
\begin{eqnarray}
(\hat \psi,\overline {B}^\pm(kU,kV))
&=&\int\limits^{\infty}_{-\infty}  \hat \psi (\omega)^*  
            B^\pm_\omega (kU,kV)~ d\omega \\
&\equiv &
\int\limits^{\infty}_{-\infty} \psi (\theta)^*
       \frac{1}{ \sqrt{2\pi} }e^{\mp ik(Ve^\theta +Ue^{-\theta})/2}  d\theta
          \quad \nonumber 
\end{eqnarray}
is a convergent integral which depends continuously on the parameters
$U$ and $V$.  Thus one says that $\overline B^\pm(kU,kV)$ is
continuous also \emph{relative} to the space of smooth test functions
of compact support.  However, it is obviously not smooth relative to
this space of test functions.

\section{THE GLOBAL REPRESENTATION}

Let us summarize the construction of the two families of generalized
functions $B^+(kU,kV)$ and $B^-(kU,kV)$, including their unique
extensions onto the event horizon:

Consider the Fourier transform representation of the boost-invariant
Cauchy evolution, Eq.(\ref{eq:M-B mode}),
\be
B^\pm_\omega (kU,kV)
=\lim_{a \to \infty}{1\over {2\pi}} \int\limits^{a}_{-a}
       e^{\mp ik(Ve^\theta +Ue^{-\theta})/2}  e^{i\omega\theta} d\theta 
\quad .
\label{eq:formula}
\ee
This integral is always well-defined, provided one remembers to
include into its definition the requirement that one take the limit
{\it after} one has done the requisite $\omega$-integration,
\begin{eqnarray}
\lefteqn{
  \int\limits^{\infty}_{-\infty} \hat\phi(\omega)^*
                                B^\pm_\omega (kU,kV)    ~d\omega =}\\
  & & =\lim_{a \to \infty}  \int\limits^{\infty}_{-\infty} \hat\phi(\omega)^*
\left[
     {1\over {2\pi}} \int\limits^{a}_{-a}
       e^{\mp ik(Ve^\theta +Ue^{-\theta})/2}  e^{i\omega\theta} d\theta 
\right]~d\omega
\quad .
\label{eq:distribution value}
\end{eqnarray}
(Here, as in example 1 of Section IV.D, one can take 
$\hat\phi(\omega)\in Z_0$, i.e. is any test function whose Fourier transform
has compact support in the $\theta$-domain)

If one keeps this rule in mind, then Eq.(\ref{eq:formula}) coincides
with the shorthand formula, Eq.(\ref{eq:M-B mode}),
\begin{equation}
B^\pm_\omega (kU,kV)={1\over {2\pi}} \int\limits^{\infty}_{-\infty}
       e^{\mp ik(Ve^\theta +Ue^{-\theta})/2}  e^{i\omega\theta} d\theta 
\quad ,
\label{eq:M-B mode again}
\end{equation}
not only whenever $(U,V)$ lies in the interior of any of the four
Rindler sectors, but also whenever $UV=0$, i.e. when $(U,V)$ lies on
any one of the event horizons.

As a point of curiosity, 
we note that in the first case, when $UV\ne 0$, $B^\pm (kU,kV)$ is a \emph{regular}
distribution, which means that the integral in Eq.(\ref{eq:M-B mode
again}) converges to an ordinary function of $\omega$. In the second
case, when $UV=0$, $B^\pm (kU,kV)$ is a \emph{singular} distribution,
which means that the integral in Eq.(\ref{eq:M-B mode again}) does not
converge to an ordinary function of $\omega$. In that case one must
resort to Eq.(\ref{eq:distribution value}) in order to impart meaning to
Eq.(\ref{eq:M-B mode again})

\section{QUANTUM MECHANICAL TUTORIAL: DISTRIBUTIONS AS SUMS OVER 
INTERFERING ALTERNATIVES}

The ($U,V$)-parametrized family of distributions $B^\pm_\omega(kU,kV)$
constitutes a flexible supply for the constructing other physically
important distributions. The transition amplitude of an accelerated
quantum system (``detector'') is a key example. Within first order
perturbation theory this amplitude, we recall, is a linear
superposition, in fact a spacetime integral, over that set of events
$(U,V)$ where the interaction between the detector and the ambient
field is non-zero. The resultant superposition (``sum over interfering
alternatives'') is a new distribution, the transition amplitude.

Of particular significance is the amplitude for the detector to make a
transition and thereby create a photon from the Minkowski vacuum.
This amplitude, which is based on the interaction action
\be
S_{int}=\int\limits^{\infty}_{-\infty}
\psi (x^\mu (\tau))~m(\tau)c(\tau)~d\tau~~,
\label{eq:interaction action}
\ee
is
\begin{eqnarray}
A^{Mink}(\omega,\tau)
&=&~_M\langle 1_\omega \vert \otimes \langle E_{fin} \vert~
\psi (x^\mu (\tau))~m(\tau)c(\tau)~
\vert 0\rangle_M\otimes \vert E_{init}\rangle \nonumber \\
&=& ~_M\langle 1_\omega \vert \psi (x^\mu (\tau))\vert 0\rangle_M~
\langle E_{fin} \vert m(\tau)c(\tau)\vert E_{init} \rangle ~~.
\label{eq:Mink amplitude}
\end{eqnarray}
Here 
\[
\{x^\mu (\tau) \}=\{U(\tau),V(\tau):~-\infty <\tau<\infty\}
\]
is the detector world line. The field is 
\be
\psi (x^\mu)=\int\limits^{\infty}_{-\infty} \{ a_\omega B^+_\omega(kU,kV)
+a_\omega^* \left[ B^+_\omega(kU,kV)\right]^*\}~d\omega~~~.
\label{eq:field}
\ee
The detector is coupled to this field by means of its dynamical property
\be
m(\tau)=e^{iH_0 \tau} m(0)e^{-iH_0 \tau}
\label{eq:monopole}
\ee
whose evolution is controlled by the detector Hamiltonian $H_0$. The
scalar $c(\tau)$ is a time-dependent coupling constant, a square
integrable function, which expresses the finite time interval, say
$-T\le \tau \le T$, during which the detector interacts with the
field. For the sake of concreteness one may assume that this coupling
function decays exponentially outside this interval. Thus one can
take $c(\tau)$ to be 
\be
c(\tau)=c_0 \left\{
\begin{array}{lc}
e^{\alpha(\tau+T)}& \tau<-T \\
1                 & -T\le \tau \le T \\
e^{-\alpha(\tau-T)}& T< \tau
\end{array}
        \right.
\label{eq:coupling function}
\ee 
Note that this function is continuous, but that it becomes
discontinuous in the limit $\alpha \to \infty$. 

\subsection{Interfering and  Exclusive Alternatives}

We say that the expression $A^{Mink}(\omega,\tau)$ is the amplitude
for finding the interacting detector-field system in the state
$~_M\langle 1_\omega \vert \otimes \langle E_{fin} ~\vert $ if, as a
result of the interaction at event $\{x^\mu (\tau) \}$, the system is
in the state $\vert \psi (x^\mu (\tau))\vert 0\rangle_M\otimes \vert
m(\tau)c(\tau)\vert E_{init}\rangle$.

Thus, for each event along the detector's world line, there is a unique
amplitude, Eq.(\ref{eq:Mink amplitude}), that (i) the detector has
made a transition to the state $\langle E_{fin} \vert$ and that (ii) a
Minkowski photon of Rindler frequency $\omega$ ($-\infty <\omega <\infty$)
has been created from the Minkowski vacuum state. This amplitude,
Eq.(\ref{eq:Mink amplitude}), is a distribution for each value of $\tau$,
and it is given with the help of Eqs.(\ref{eq:field}) and (\ref{eq:monopole})
by
\[
A^{Mink}(\omega,\tau)=\langle E_{fin} \vert m(0)\vert E_{init} \rangle
e^{iE_{fin} \tau} c(\tau) \left[ B^+_\omega(kU(\tau),kV(\tau))\right]^*
                                        e^{-iE_{init} \tau}
\]
The total amplitude is obtained by summing over all these interfering
alternatives labeled by the events ($-\infty<\tau<\infty$)
on the detector world line:
\begin{eqnarray}
\int\limits^{\infty}_{-\infty} A^{Mink}(\omega,\tau)~d\tau
&=& ~_M\langle 1_\omega \vert \otimes \langle E_{fin} ~\vert S_{int}~
        \vert 0\rangle_M \otimes \vert E_{init} \rangle\nonumber \\
&=& \langle E_{fin} \vert m(0)\vert E_{init} \rangle
\int\limits^{\infty}_{-\infty} 
e^{iE_{fin} \tau} c(\tau) \left[ B^+_\omega(kU(\tau),kV(\tau))\right]^*
                                e^{-iE_{init} \tau} ~d\tau \nonumber \\
&\equiv& A^{Mink~Emit}(\omega)~~.
\label{eq:total amplitude}
\end{eqnarray}
This is another distribution, the complex amplitude that the detector
make a transition to the state $\langle E_{fin} \vert$ while
simultaneously creating a Minkowski photon of Rindler frequency
$\omega$ ($-\infty<\omega<\infty$) during the time interval 
imposed by the function $c(\tau)$. The following two points need
to be reemphasized:
\begin{enumerate}
\item
The totalized transition amplitude $A^{Mink~Emit}(\omega)$ is quite
general: no assumptions about the detector world line have been made
as yet.
\item
This amplitude is for the emission of a Minkowski photon whose
\emph{moment of energy} 
\cite{MTW5}
relative to the reference event $(t_0,z_0)$ is $\omega$ 
($-\infty<\omega<\infty$). Even though we have been referring to this
quantity as the ``Rindler frequency'' of a Minkowski photon,
one should not confuse this property with the ``Rindler frequency''
of a Rindler-Fulling quantum. The first generates Lorentz boosts
in all of Minkowski spacetime, while the second generates 
Rindler time translations in Rindler Sector $I$ (or $II$) only.
\end{enumerate}

The emission of Minkowski photons having this, that, or the other
Rindler frequency $\omega$ ($-\infty<\omega<\infty$) when the detector
makes a transition are not \emph{interfering alternatives},
but instead are \emph{exclusive alternatives} \cite{Feynman1}.
This is why the unconditional detector transition probability
is the sum, or more precisely the integral, of the \emph{squared} modulus
of $A^{Mink~Emit}(\omega)$, the probability amplitude:
\be
\int\limits^{\infty}_{-\infty} 
\vert A^{Mink~Emit}(\omega)\vert^2~d\omega~~~.
\label{eq:M-probability}
\ee
This unconditional transition probability is the total 
probability for the detector to make a transition as
predicted by first order perturbation theory based on 
Eq.(\ref{eq:interaction action}).

Quantum mechanics tells us that
$\vert A^{Mink~Emit}(\omega)\vert^2~d\omega$ is the differential
detector transition probability conditioned by the emission of a 
Minkowski photon of type $\omega$ ($-\infty<\omega<\infty$).
Consequently,
nature demands that the amplitude $A^{Mink~Emit}(\omega)$ be a very
special type of distribution, namely a square-integrable function of
$\omega$.
Thus our formulation must satisfy
\be
\int\limits^{\infty}_{-\infty} 
\vert A^{Mink~Emit}(\omega)\vert^2~d\omega\equiv {\mathcal{P}}^{Mink~Emit}<\infty~~~.
\ee
Higuchi, Matsas, and Peres
\cite{Higuchi} (HMP) have shown that this is indeed the case
if and only if the detector is switched 
on and off in a continuous way (i.e. if and only if 
$c(\tau)$ depends continuously on $\tau$) whenever it
undergoes uniform acceleration in Rindler Sector $I$.

\subsection{Relations Among Transition Amplitudes and Probabilities}

Let us assume, therefore, that the interaction of the uniformly
accelerated detector is confined to that portion of its history which
lies strictly in Rindler Sector $I$.  In that case the amplitude
$A^{Mink~Emit}(\omega)$, Eq.(\ref{eq:total amplitude}), is a regular
distribution, an ordinary function of $\omega$ whose form for
$\omega>0$ is
\begin{eqnarray}
A^{Mink~Emit}(\omega)
&=& \langle E_{fin} \vert m(0)\vert E_{init} \rangle
\int\limits^{\infty}_{-\infty} 
e^{iE_{fin} \tau} c(\tau) 
                        \left[ 
\frac{\sqrt{2\sinh \pi \omega}}{\pi} K_{i\omega}(k\xi)e^{-i\omega\tau}~~
                        \right]^*
\frac{e^{\pi\omega /2}}{\sqrt{2\sinh \pi \omega}}
e^{-iE_{init} \tau} ~d\tau \nonumber \\
&=& \langle E_{fin} \vert m(0)\vert E_{init} \rangle 
\int\limits^{\infty}_{-\infty} e^{iE_{fin} \tau} c(\tau) 
~_R\langle 1_\omega \vert ~\psi (x^\mu (\tau))~\vert 0\rangle_R~~
e^{-iE_{init} \tau} ~d\tau
\left(1+\frac{1}{\exp 2\pi\omega -1} \right)^{1/2} \nonumber\\
&\equiv&
A^{Rind~Emit}(\omega) \left(1+\frac{1}{\exp 2\pi\omega -1} \right)^{1/2}~~
~~~~~~~~~~0<\omega
\label{eq:R-emission amplitude}
\end{eqnarray}
Here $A^{Rind~Emit}(\omega)$ is the amplitude for the process of the
detector making the transition $\vert E_{init}\rangle \to \vert
E_{fin}\rangle$ while emitting a Fulling-Rindler quantum of 
Rindler frequency $\omega>0$ into the Fulling-Rindler vacuum
$\vert 0\rangle_R$. One is forced into this identification of
$A^{Rind~Emit}(\omega)$ because the normal mode expansion of
the Klein-Gordon field in Rindler Sector $I$ is
\[
\psi (x^\mu)=\int_0^\infty
\{ A_\omega \frac{\sqrt{2\sinh \pi \omega}}{\pi} 
                                        K_{i\omega}(k\xi)e^{-i\omega\tau}~~
+ A_\omega^* \frac{\sqrt{2\sinh \pi \omega}}{\pi} 
                                        K_{i\omega}(k\xi)e^{i\omega\tau}
                        ~\}~d\omega
\]
with
\[
A_\omega \vert0\rangle_R=0~~~.
\]
However, when $\omega=-\vert\omega\vert <0$, the amplitude 
$A^{Mink~Emit}(\omega)$ leads to a different quantity:
\begin{eqnarray}
A^{Mink~Emit}(-\vert\omega\vert)
&=& \langle E_{fin} \vert m(0)\vert E_{init} \rangle
\int\limits^{\infty}_{-\infty} 
e^{iE_{fin} \tau} c(\tau) 
                        \left[ 
\frac{\sqrt{2\sinh \pi \vert \omega\vert}}{\pi} K_{i\omega}(k\xi)e^{i\vert\omega\vert\tau}~~
                        \right]^*
\frac{e^{-\pi\vert\omega\vert /2}}{\sqrt{2\sinh \pi \vert \omega\vert}}
e^{-iE_{init} \tau} ~d\tau \nonumber \\
&=& \langle E_{fin} \vert m(0)\vert E_{init} \rangle 
\int\limits^{\infty}_{-\infty} e^{iE_{fin} \tau} c(\tau) 
~_R\langle 0 \vert ~\psi (x^\mu (\tau))~\vert 1_{\vert\omega\vert}\rangle_R~~
e^{-iE_{init} \tau} ~d\tau
\left(\frac{1}{\exp 2\pi\vert \omega\vert -1} \right)^{1/2}\nonumber\\
&\equiv&
A^{Rind~Abs}(\vert\omega\vert) \left(\frac{1}{\exp 2\pi\vert \omega\vert -1} \right)^{1/2}~~
~~~~~~~~~~\omega <0
\label{eq:R-absorption amplitude}
\end{eqnarray}
In this case $A^{Rind~Abs}(\vert\omega\vert)$ is the amplitude for the
process of the detector making the same transition $\vert
E_{init}\rangle \to \vert E_{fin}\rangle$, but this time absorbing a
Fulling-Rindler quantum (of frequency $\vert\omega\vert=-\omega>0$)
and leaving the field in the Fulling-Rindler vacuum $\vert0\rangle_R$.
The relationships, Eqs.(\ref{eq:R-emission amplitude}) and
(\ref{eq:R-absorption amplitude}), between the probability amplitudes
imply that
\begin{equation}
\int\limits^{\infty}_{-\infty} 
                        \vert A^{Mink~Emit}(\omega)\vert^2~d\omega
=\int\limits^{\infty}_0 
        \vert A^{Rind~Emit}(\omega)\vert^2~
\left(1+\frac{1}{\exp 2\pi\omega -1} \right)~d\omega
+
\int\limits^{\infty}_0 
        \vert A^{Rind~Abs}(\omega)\vert^2~
\left(\frac{1}{\exp 2\pi\omega -1} \right)~d\omega~~~,
\label{eq:basic equation}
\end{equation}
or more compactly
\be
{\mathcal{P}}^{Mink~Emit}={\mathcal{P}}^{Rind~Emit}+{\mathcal{P}}^{Rind~Abs}~~.
\label{eq:basic equation2}
\ee
This equation relates the detector transition probability when a Minkowski
photon is emitted to the transition probability when a Fulling-Rindler quantum
is emitted \emph{or} absorbed. 

\subsection{Finite-time Detector}

What is of particular interest, is how these 
probabilities depend on (i) the time duration, $2T$, of the interaction
strength, Eq.(\ref{eq:coupling function}), between the detector and the 
ambient field, (ii) the detector's switching time $\frac{1}{\alpha}$,
and (iii) the detector's transition energy $\Delta E=E_{fin}-E_{init}$.
Quantum mechanics answers these questions by furnishing us with
explicit expressions for the three probability amplitudes,
\begin{eqnarray}
A^{Mink~Emit}(\omega)
&=& \langle E_{fin} \vert m(0)\vert E_{init} \rangle
f(\omega-\Delta E) \frac{e^{\pi\omega /2}}{\pi} K_{i\omega}(k\xi )
                                        ~~~~~-\infty<\omega<\infty 
                                        \label{eq:M-emission amplitude}\\
A^{Rind~\begin{array}{c}
                Emit\\
                Abs
        \end{array} }(\omega)
&=& \langle E_{fin} \vert m(0)\vert E_{init} \rangle
f(\omega\mp\Delta E) \frac{\sqrt{2\sinh\pi\omega}}{\pi} K_{i\omega}(k\xi )
                                                        ~~~~~0<\omega<\infty ~~~,
\end{eqnarray}
where 
the function
\begin{eqnarray}
f(\omega \mp\Delta E)
&\equiv&
        \int\limits^{\infty}_{-\infty} 
        e^{iE_{fin} \tau} c(\tau) e^{\mp i\omega\tau} e^{-iE_{init} \tau} ~d\tau \nonumber \\
&=&
\frac{2\alpha}{\alpha^2 +(\omega \mp \Delta E)^2} \cos T(\omega \mp \Delta E)
+2\left( \frac{1}{\omega \mp \Delta E}
                -\frac{\omega \mp \Delta E}{\alpha^2 +(\omega \mp \Delta E)^2}
        + \right) 
                                \sin T(\omega \mp \Delta E)
\end{eqnarray}
is a temporal overlap integral. It is an even function of its
argument, and it expresses the resonance between the Rindler field
oscillator $e^{\mp i\omega\tau}$ and the transition frequency
($E_{fin}-E_{init}=\Delta E$) oscillator $e^{-i\Delta E\tau}$ of the
detector.

By introducing the probability amplitudes $A^{Rind~Emit}$ and 
$A^{Rind~Abs}$ into Eq.(\ref{eq:basic equation2}), which HMP infer from 
Unruh's original work, they show that probability is finite, i.e.
\be
{\mathcal{P}}^{Rind~Emit}+{\mathcal{P}}^{Rind~Abs}
={\mathcal{P}}^{Mink~Emit}<\infty~~,
\label{eq:basic equation3}
\ee
if and only if the function $c(\tau)$ is continuous, i.e. the
switching time $\frac{1}{\alpha}$ is non-zero.
Using their result we conclude from this that, under these conditions,
the amplitude $A^{Mink~Emit}$, Eq.(\ref{eq:M-emission amplitude}) is a
distribution which is a square-integrable function of $\omega$. 

The physical significance of HMP's finiteness result is that the total
probability in Eq.(\ref{eq:basic equation3}) complies with Fermi's
golden rule by being proportional to the finite time, $\Delta\tau=2T$,
that the detector is switched on. 

It is a fact that physical detectors can be accelerated linearly and
uniformly only for a finite amount of time. Consequently, we can take the
switch-on time, $\Delta\tau=2T$, to be the finite amount of time
that the detector is being accelerated.

In the next section (VII.E) we introduce a dynamical wave complex
(``acceleron'') which also has a finite life time,
\[
\Delta \tau=\varepsilon ~~~.
\]
The existence, identity, and detailed physical properties of this wave
complex can be predicted by calculating the interaction between it and 
a finite-time detector. The interaction between the two evokes a maximum
response in the detector if the life times of the two match,
\[
\varepsilon \approx 2T~~.
\]
Suppose the detector is in resonance with this wave complex, i.e.
\[
E_{fin} -E_{init} \approx \overline\omega ~~~,
\]
where $\overline\omega$ is its characteristic frequency defined in
Eq.(\ref{eq:slabthickness}. Then these two systems exchange ``boost'' energy
periodically in a manner analogous to two interacting pendulums. Such
a periodic energy exchange would presumably be accessible to
experiments. This resonance interaction would yield enough information
about the state of this wave complex to allow it being used as a new
way of probing the properties of spacetime. This detection process
corresponds to using photons to watch planets and comets to obtain
information about their positions and velocities for the purpose of
probing the properties of spacetime (gravitation).
Space limitations forbid us to pursue this line of inquiries
in this article.

\section{WAVELETS AND WAVE PACKET HISTORIES}

The ($U,V$)-parametrized family of Minkowski-Bessel distributions
constitutes a linear one-to-one transform from the space spanned by
orthonormal wavelets to the space of wave packet histories.

A wavelet is a square-integrable function which is localized in phase
space, i.e. the function $\phi(\theta)$ and its Fourier transform
$\hat \phi (\omega )$ are concentrated on small sets.  We shall use
the term ``wavelet'' in a sense which is (slightly) more inclusive
than that used by the ``wavelet theory'' community
\cite{IEEE}-\cite{I.Daubechies}. Our orthonormal
``wavelets'' agree with theirs in that ``wavelets'' are localized in
the given domain (say, $-\infty<\theta<\infty$) and the Fourier domain
($-\infty<\omega<\infty$).  However, workers in ``wavelet theory''
restrict the term orthonormal ``wavelets'' to sequences of functions
whose localization in the given domain is described by 
compression and translation operations applied to a single ``mother
wavelet'' function 
\cite{Cohen and Kovacevic},\cite{Hess-Nielsen and Wickerhauser}.
As a consequence, this restricts their wavelets 
to those whose adjacent frequencies in a (wavelet) Fourier series differ
from each other by an octave in the Fourier domain. We, on the other hand,
include also those which realize the windowed Fourier transform, such
as Eqs.(\ref{eq:windowed exponential}) and (\ref{eq:Fourier windowed
exponential}) in the ensuing mathematical tutorial.

Either of the wavelets,
$\phi(\theta)$ or $\hat \phi (\omega )$, determines a unique wave packet history,
a solution to the Klein-Gordon equation:
\begin{eqnarray}
\phi^\pm (kU,kV)&=&\int_{-\infty}^\infty \hat \phi (\omega )^*
   B^\pm_\omega (kU,kV)~d\omega 
                                        \label{eq:omegawave transform}\\
                &=&\int_{-\infty}^\infty \phi(\theta)^*\frac{1}{ \sqrt{2\pi} }
        e^{\mp ik(Ve^\theta +Ue^{-\theta})/2}~d\theta~~.
\label{eq:thetawave transform}
\end{eqnarray}
As noted in the opening sentence, the correspondence between the set
of wave packet histories and the set of square-integrable functions is
linear and one-to-one. In fact, the inverse transformation is
\begin{eqnarray}
\hat \phi (\omega)^*&=&\pm \langle B^\pm_\omega ,\phi^\pm \rangle \\
\phi(\theta)^*  &=&\pm \langle \frac{1}{ \sqrt{2\pi} }e^{\mp ik(Ve^\theta +Ue^{-\theta})/2},\phi^\pm \rangle
~~~.
\end{eqnarray}
This follows from the Klein-Gordon orthonormality of the Minkowski-Bessel
and the plane wave modes:
\begin{eqnarray}
\langle B^\pm_\omega ,B^\pm_{\omega'}\rangle 
&\equiv &
    \frac{i}{2} \int_{-\infty}^\infty \left[ (B^\pm_\omega)^*
    {{\partial} \over\partial t} B^\pm_{\omega'}-
        {{\partial} \over\partial t}(B^\pm_\omega)^*B^\pm_{\omega'}
                        \right]~dz      \label{eq:Klein-Gordon inner product} \\
&=      & \pm \delta (\omega -\omega') \\
\langle B^+_\omega ,B^-_{\omega'}\rangle &=& 0 \nonumber\\
\langle \frac{e^{\mp ik(Ve^\theta +Ue^{-\theta})/2} }{\sqrt{2\pi}},
        \frac{e^{\mp ik(Ve^{\theta'} +Ue^{-\theta'})/2} }{\sqrt{2\pi}}\rangle 
&=&\pm \delta (\theta-\theta')~~~. \nonumber
\end{eqnarray}

\subsection{The Phase Space}

The outstanding virtue of the two transforms, Eqs.(\ref{eq:omegawave
transform}) and (\ref{eq:thetawave transform}) is the Fourier duality
of the two sets, $\{ \phi\}$ and $\{ \hat \phi\}$, of
square-integrable functions. They serve to represent the Klein-Gordon
solutions in two different and mutually exclusive ways: one relative
to an inertial frame via plane wave modes, the other relative to the
four Rindler quadrants of a pair of accelerated frames via
Minkowski-Bessel modes.

This Fourier duality implies the existence of the familiar
two-dimensional phase space. As demonstrated in the ensuing tutorial, 
this space is partitioned by a
chosen basis of orthonormal wavelets into equal-area phase space
cells, each one of area $2\pi$. 
\begin{figure}[ht!]
\epsfclipon
\epsffile[-125 -20 283 268]{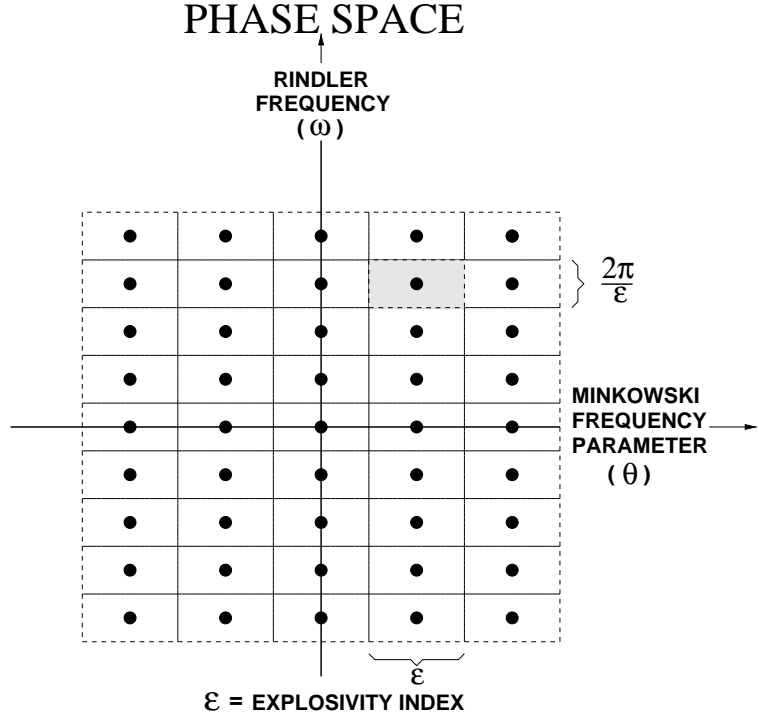}
       \epsfverbosetrue
\caption{
Two-dimensional phase space spanned by the Minkowski frequency
parameter $\theta$ ( Eq.(\ref{eq:Minkowski frequency}) ) and by the
Rindler frequency $\omega$.  The windowed Fourier basis of orthonormal
wavelets ( Eqs.(\ref{eq:windowed exponential})-(\ref{eq:Fourier
windowed exponential}) ) partitions this space into phase space cells
of equal area $2\pi$. The support of each wavelet and its Fourier
transform are concentrated horizontally and vertically within each
dashed rectangle. A heavy dot locates the amplitude maximum of the
corresponding wavelet and its Fourier transform. The shape of the area
elements is controlled by the freely adjustable parameter
$\varepsilon$. Physically it indicates the rate at which each Klein-Gordon
wave packet collapses and explodes.}
\label{fig:cartesian phasespace}
\end{figure}
This set of wavelets
and their Fourier transforms,
\begin{equation}
\{\phi_{j\ell}^\varepsilon(\theta)~:~~j,\ell=0,\pm 1,\pm 2, \cdots\}
\end{equation}
and
\begin{equation}
\{\hat \phi_{j\ell}^\varepsilon(\omega)~:~~j,\ell=0,\pm 1,\pm 2, \cdots\}
\end{equation}
satisfy the orthonormality condition
\begin {equation}
\int_{-\infty}^\infty [\phi_{j\ell}^\varepsilon(\theta)]^*\phi_{j'\ell'}(\theta)]~d\theta=
\int_{-\infty}^\infty [\hat \phi_{j\ell}^\varepsilon(\omega)]^*\hat \phi_{j'\ell'}(\omega)~d\omega=
\delta_{jj'}\delta_{l\ell'}~~~,
\end{equation}
and, as pointed out in the ensuing tutorial, they form a complete set.
However, its most interesting feature from the viewpoint of physics is
the freedom one has in choosing the size of the domain neighborhood
on which the wavelet is concentrated. For the wavelet, Eq.(\ref{eq:windowed
exponential}),
\begin{equation}
\phi^\varepsilon_{j\ell}(\theta)~~~\textrm{this~neighborhood~is}~~~
(\ell - \frac{1}{2})\varepsilon \le \theta \le (\ell + \frac{1}{2})\varepsilon ,
\end{equation}
while for its Fourier transform, Eq.(\ref{eq:Fourier windowed
exponential}),
\begin{equation}
\hat \phi^\varepsilon_{j\ell}(\omega)~~~\textrm{this~neighborhood~is}~~~
(j - \frac{1}{2})\frac{2\pi}{\varepsilon} \le \omega \le 
              (j + \frac{1}{2})\frac{2\pi}{\varepsilon}.
\end{equation}
The size of these neighborhoods is obviously controlled by the
positive parameter $0<\varepsilon <\infty$. 

There are two extreme
cases which have physically important consequences:

\noindent
1.  Whenever $\varepsilon\ll 1$, the wavelet basis elements
     $\phi_{j\ell}^\varepsilon(\theta)$ are concentrated in a very small
     subinterval in the $\theta$-domain. This implies, of course, that
     their Fourier transforms, $\hat \phi_{j\ell}^\varepsilon(\omega)$,
     are very spread out in the $\omega$-domain.

\noindent
2.  Whenever $\varepsilon\gg 1$, the wavelet basis elements
     $\hat \phi_{j\ell}^\varepsilon(\omega)$ are concentrated in a very small
     subinterval in the $\omega$-domain. In that case we know that
     their inverse Fourier transforms, $\phi_{j\ell}^\varepsilon(\theta)$,
     are correspondingly spread out in the $\theta$-domain.

\noindent
Thus by varying the parameter $0<\varepsilon<\infty$ one deforms
continuously the wavelets $\phi_{j\ell}^\varepsilon(\theta)$ from
ones which are highly localized in the $\theta$-domain to those whose
Fourier transforms, $\hat \phi_{j\ell}^\varepsilon(\omega)$, are highly
localized in the $\omega$-domain. In fact, in the asymptotic limit
this localized behavior is proportional to a Dirac delta function.

\begin{figure}[ht!]
\epsfclipon
\epsffile[-75 -20 392 206]{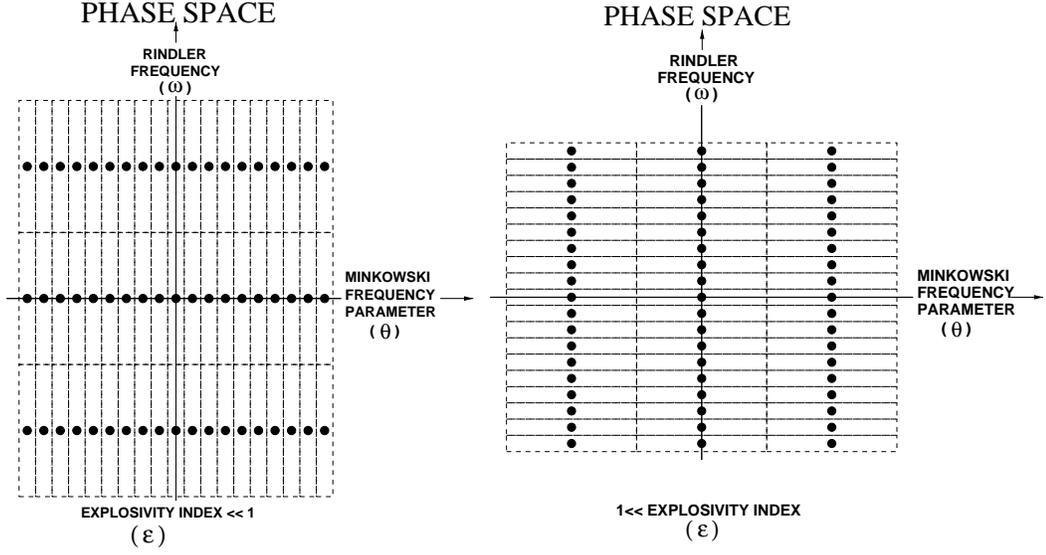}
       \epsfverbosetrue
\caption{ Two different partitionings of phase space. The elementary
phase space areas have the same magnitude, but their shapes are
different. They are characterized by the explosivity index
$\varepsilon$ defined in Figure \ref{fig:cartesian phasespace}. The tall
and skinny elements ($\varepsilon \ll 1$) define Klein-Gordon wave
packet histories with well-defined mean Minkowski frequency but
indeterminate Rindler frequency. Such wave packets \emph{contract and
re-expand non-relativistically} in their respective frames of
reference. By contrast, the short and squatty elements ($1 \ll
\varepsilon $) define Klein-Gordon wave packet histories with
well-defined mean Rindler frequency but indeterminate Minkowski
frequency. Such wave packets \emph{collapse and re-explode
relativistically} in their respective frames of reference.  }
\label{fig:cells_skinny_and_squatty}
\end{figure}

This process of deformation alters the shape of the phase space cells,
whose union makes up the phase space. Assume, as is done in the ensuing
tutorial, that this two-dimensional
phase space is coordinatized horizontally by $-\infty <\theta <\infty$
and vertically by $-\infty <\omega <\infty$.  Then, as depicted in
Figure~\ref{fig:cells_skinny_and_squatty}, this deformation corresponds to changing the shape of each
phase space area element from being tall and skinny ($\varepsilon\ll
1$) to being short and squatty ($\varepsilon\gg 1$).

\subsection{Complete Set of Smooth Orthonormal Wave Packet Histories}

Every complete set of orthonormal wavelets determines a corresponding set of
Klein-Gordon orthonormal wave packet histories.

Consider the orthonormal wavelets
\be
\phi_{j\ell}^\varepsilon (\theta)= \frac{1}{\sqrt{\varepsilon}} 
\exp \left(-i\frac{2\pi}{\varepsilon}j\theta \right)\times \left\{
\begin{array}{ccc}
1& ~~\textrm{whenever}& (\ell-\frac{1}{2})\varepsilon \le \theta \le (\ell+\frac{1}{2})\varepsilon\\
0& ~~\textrm{whenever}& \frac{\varepsilon}{2}<\vert \ell\varepsilon 
                                                            -\theta \vert
\end{array}
\right\} ~~~j,\ell=0,\pm1,\pm,2,\cdots~~,
\label{eq:second windowed exponential}
\ee
and their Fourier transforms
\begin{eqnarray}
\hat \phi_{j\ell}^\varepsilon (\omega)
&=&
\frac{1}{\sqrt {2\pi \varepsilon}}
\int _{(\ell-{1\over 2})\epsilon }^{(\ell+{1\over 2}) \varepsilon } 
             \exp {i\left( \omega - {2\pi \over \varepsilon}j\right) \theta }
     ~d\theta 
\quad ~~~~~~~~~~~~~~~~~~~~~~~~~~~~~~~j,\ell=0,\pm 1,\cdots \nonumber\\
&=&
\frac{1}{\sqrt {2\pi \varepsilon}}
\exp {i\left(  \omega- {2\pi \over \varepsilon}j \right) l\epsilon } \times
   \frac{2 \sin \left(  \omega- {2\pi \over \epsilon}j\right) 
          \frac{\epsilon}{2} }{\left(  \omega-{2\pi \over \epsilon}j \right)},
\label{eq:second Fourier windowed exponential}
\end{eqnarray}
These wavelets are taken from the ensuing tutorial, Section VIII.B.
The corresponding wave packet histories,
\begin{eqnarray}
\phi^{\varepsilon \pm}_{j\ell}(kU,kV)
        &\equiv&\int_{-\infty}^\infty   \hat \phi^\varepsilon_{j\ell} (\omega)^*
        B^\pm_\omega(kU,kV) ~d\omega  \label{eq:linear map omega}\\
        &=&\int_{-\infty}^\infty \phi^\varepsilon_{j\ell} (\theta)^*
\frac{1}{ \sqrt{2\pi} }e^{\mp ik(Ve^\theta +Ue^{-\theta})/2}  d\theta
~~~j,\ell =0,\pm1,\pm2,\cdots~~~, \label{eq:linear map theta}
\end{eqnarray}
are Klein-Gordon-orthonormal,
and the correspondence itself is obviously one-to-one and linear.
Furthermore, the completeness of the set of orthonormal wavelets, i.e.
\begin{eqnarray}
\sum_{j=-\infty}^\infty \sum_{\ell=-\infty}^\infty
\phi^\varepsilon_{j\ell} (\theta)~\phi^\varepsilon_{j\ell} (\theta')^*
&=&\delta(\theta-\theta') \\
\sum_{j=-\infty}^\infty \sum_{\ell=-\infty}^\infty
\hat \phi^\varepsilon_{j\ell} (\omega)~\hat \phi^\varepsilon_{j\ell} (\omega')^*
&=&\delta(\omega-\omega')
\end{eqnarray}
implies the completeness of the corresponding set wave packet histories.
This means that 
\begin{em}
any 
solution to the Klein-Gordon Eq.(\ref{eq:1}), say $\psi(kU,kV)$, can 
be expressed as a linear combination of these histories:
\end{em}
\be
\psi (kU,kV)=\sum_{j=-\infty}^\infty \sum_{\ell=-\infty}^\infty
a_{j\ell}\phi^{\varepsilon+}_{j\ell}(kU,kV) ~ 
                                + ~b^*_{j\ell}\phi^{\varepsilon-}_{j\ell}(kU,kV)
\label{eq:scalar solution}
\ee
where the coefficients are expressed in terms of the Klein-Gordon
inner product, Eq.(\ref{eq:Klein-Gordon inner product}),
\be
a_{j\ell}=\langle \phi^{\varepsilon+}_{j\ell}, \psi  \rangle \quad ; \quad 
b^*_{j\ell}=- \langle \phi^{\varepsilon-}_{j\ell},\psi  \rangle~~~.
\ee

The most perspicuous feature about the wave packet histories is their
dependence on the common parameter $0<\varepsilon <\infty$.  When
$\varepsilon \ll 1$, and the wavelet $\phi_{j\ell}^\varepsilon
(\theta)$ is highly localized in the Minkowski frequency ($\theta$)
domain, then the wave packet history $\phi^{\varepsilon
  \pm}_{j\ell}(kU,kV)$ is a long, but finite, wave train travelling with
a spacetime velocity whose direction is determined by the localized
$\theta$-value of the wavelet. As is well known and is shown in the
next subsection, the size of this wave train is not constant. In fact, it
contracts, reaches a minimum size, and then re-expands at a rate whose
asymptotic value is
\[
\frac{v}{c}=\varepsilon~~~~\ll~1
\]
in the proper rest frame of the wavetrain packet. This process is very
gentle, with an appearance very much like that of a diffusive evolution.

When $1\ll \varepsilon$, and the wavelet $\hat
\phi_{j\ell}^\varepsilon (\omega)$ is highly localized in the Rindler
frequency ($\omega$) domain, then the wave packet history
$\phi^{\varepsilon \pm}_{j\ell}(kU,kV)$ manifests itself in a violent way
relative to a globally inertial reference frame. In fact, from the
boundary of the history in Figures \ref{fig:amplitude splitting},
\ref{fig:reflected modes}, or \ref{fig:WKB modes} one sees that the
history consists of a collapse followed by an explosion.  This process
is characterized by contraction and expansion rates which are
relativistic into both the negative and the positive $z$-direction,
and their asymptotic magnitude is
\[
\frac{v}{c}=\tanh \frac{\varepsilon}{2\pi}~~~~\approx 1~~.
\]
This asymptotically relativistic expansion (or contraction) makes the
process have the appearance of an explosion (or collapse). Indeed, the
violence of this evolutionary unfolding is controlled entirely by the
positive parameter $\varepsilon$.  For this reason it is appropriate
to refer to $\varepsilon$ as the \emph{explosivity index} of the wave
complex, as is done in Figures \ref{fig:cartesian phasespace} and
\ref{fig:cells_skinny_and_squatty}.

\subsection{Two Asymptotic Limits}

A wave packet and a classical particle are similar in that they both
refer to a localized property such as mass, or charge, or
electromagnetic energy, and so on. This similarity is the means by
which one identifies the difference in their degree of localization.
Quantitatively one does this with the parameter $0<\varepsilon
<\infty$, the ``conceptual common denominator'', which distinguishes
between different kinds of wave packet evolutions. Fundamentally,
there are only two of them: the \emph{Inertial Frame Limit},
characterized by $\varepsilon \ll 1$ and the \emph{Rindler Frame Limit}
characterized by $1\ll \varepsilon$.

In the ``$\varepsilon \ll 1$'' limit each wave packet has a (well-defined)
group velocity which defines a unique inertial frame. This limit
accommodates the familiar concept of a classical particle.

By contrast, in the ``$1\ll \varepsilon$'' limit there is no such
thing as a group velocity, and hence no such thing as an inertial
frame defined by parallel world lines \cite{Marzke and Wheeler} of
classical particles in a state of free float \cite{Taylor and
  Wheeler}.  Instead, each evolving wave complex is characterized by a
Lorentzian Mach-Zehnder interference process whose amplitude
splitting, reflection, and interference defines the four Rindler
sectors of a pair of oppositely accelerated frames.

\subsubsection{Inertial Frame Limit}

One arrives at the concept of a classical particle by letting
$\varepsilon \ll 1$ and then letting the proper rest mass of a quantum
become very large. In this limiting process, a wave packet history
becomes a particle (or antiparticle) world line with a sharply defined
tangent at each sharply defined event.
Symbolically we express this circumstance by
\begin{equation}
\lim_{
  \begin{array}{c}
    \varepsilon \ll 1\\
    k \rightarrow \infty
  \end{array}
                    }\phi^{\varepsilon \pm}_{j\ell}(kU,kV)
= \textrm{``world~line of} \left\{
\begin{array}{c} 
                 \textrm{a ~particle''~~~~~(upper~sign)}\\
                 \textrm{an ~antiparticle''~(lower~sign)}
\end{array}  
\right.
\label{eq:particle limit}
\end{equation}
The three steps which comprise this process are as follows:

\noindent 1.) Keeping
\[
\frac{2\pi}{\varepsilon}j=\textrm{fixed~and~finite}
\]
and
\[
\ell \varepsilon =\textrm{fixed}~~,
\]
evaluate the integral representation, Eq.(\ref{eq:linear map theta}),
of the wave packet $\phi^{\varepsilon \pm}_{j\ell}(kU,kV)$ in the
asymptotic limit of $\varepsilon \ll 1$. The technique for doing
this is straight forward and is described in a tutorial (Section IX:
``Wave Packets via Constructive Interference''). Using it, one finds
\[
\phi^{\varepsilon \pm}_{j\ell}(kU,kV)=~A~e^{iS},~~~~~(\varepsilon \ll 1)
\]
where
\begin{equation}
e^{iS}=\exp[\mp ik(Ve^{\overline{\theta} }+Ue^{-\overline{\theta}})/2
+\frac{2\pi}{\varepsilon}j\overline{\theta}]
~~~~~\overline{\theta}=\ell\varepsilon
\label{rapidly varying phase factor}
\end{equation}
is a rapidly varying function of spacetime, while the factor $A$ is 
a gently varying Gaussian envelope of constructive interference whose 
squared modulus is
\[
\vert A \vert ^2 \propto exp \left\{
  -\varepsilon ^2 
  \left(\frac{\partial S}{\partial \overline{\theta}} \right)^2
  \left[1+\left( 
\varepsilon^2\frac{\partial^2 S}{\partial \overline{\theta}^2} \right)^2
\right]^{-1}
\right\}
\]
\noindent 2.) Use the magnitude of this factor to identify that region
of spacetime where the wavepacket amplitude is substantially different 
from zero. A necessary condition for this to be so is that the squared
\emph{constructive interference parameter}
\begin{equation}
\delta^2\equiv
  \varepsilon ^2 
  \left(\frac{\partial S}{\partial \overline{\theta}} \right)^2
  \left[1+\left( 
\varepsilon^2\frac{\partial^2 S}{\partial \overline{\theta}^2} \right)^2
\right]^{-1} 
\label{eq:theta constructive interference parameter}
\end{equation}
satisfy
\begin{equation}
\delta^2 \le 1~~.
\label{locus of events inequality}
\end{equation}
The boundary of the spacetime region of constructive interference is
characterized by $\delta^2=1$. If at an event the wave packet has
non-negligible squared modulus, then at that event the inequality,
Eq.(\ref{locus of events inequality}), must prevail. In brief, the
locus of events of non-negligible squared modulus is given by
\begin{equation}
\frac{\varepsilon^2}{\delta^2} 
\left(\frac{\partial S}{\partial \overline{\theta}} \right)^2
-\left( \varepsilon^2\frac{\partial^2 S}{\partial \overline{\theta}^2} \right)^2
=1,~~~~~\delta^2\le 1~~.
\label{region of events}
\end{equation}
The spacetime region determined by this inequality is the Lorentzian
version of what in Euclidean wave optics would be a converging, and
then a diverging, Gaussian beam\cite{Yariv}. It is important to note
that $\varepsilon \ll 1$ does \emph{not} imply that one may omit the
term $\left( \varepsilon^2\frac{\partial^2 S}{\partial
    \overline{\theta}^2} \right)^2$. Indeed,
$\varepsilon^2\frac{\partial^2 S}{\partial \overline{\theta}^2}$
becomes non-negligible in the past and in the future.  This fact is a
reflection of the fact that the ``Lorentzian beam'', like its
Euclidean analogue, never stays parallel.  The perpendicular cross section
of the beam increases without limit in the past and in the future. 

As already noted, the boundary of this
Lorentzian beam is $\delta^2=1$. With the help of the plane wave phase
in Eq.(\ref{rapidly varying phase factor}), namely
\[
S=\mp k(t-t_0)\cosh \overline \theta \mp k(z-z_0)\sinh \overline \theta
+\frac{2\pi}{\varepsilon} j \overline \theta ~~,
\]
this boundary, Eq.(\ref{region of events},) is the pair of conjugate timelike hyperbolas
\begin{equation}
\frac{(\Delta z')^2}{a^2} -\frac{t'^2}{b^2}=1 ~~,
\end{equation}
where, first of all
\begin{equation}
a=\frac{1}{k\varepsilon}
\label{eq:proper width}
\end{equation}
is the \emph{initial proper width} of the wave packet, secondly
\begin{equation}
\frac{a}{b}=\varepsilon
\label{eq:proper expansion rate}
\end{equation}
is its \emph{proper (asymptotic) expansion rate}, and finally
\begin{equation}
\Delta z'=(z-z_0)\cosh \overline \theta +(t-t_0)\sinh \overline \theta
\mp \frac{2\pi}{k\varepsilon}j;~~~~~\overline \theta=\ell \varepsilon
\label{eq:proper halfwidth}
\end{equation}
is its \emph{proper halfwidth} after an \emph{elapsed proper time}
\[
t'=(t-t_0)\cosh \overline \theta +(z-z_0)\sinh \overline \theta~~.
\]
\noindent 3.) Taking note of the fact that the squared half width
of a wave packet is
\begin{equation}
(\Delta z')^2=\frac{1}{\varepsilon ^2 k^2} +\varepsilon ^2 t'^2~~,
\end{equation}
consider the problem of finding \emph{that} wave packet, and its history,
for which 
\begin{equation}
(\Delta z')^2 =~minimum
\end{equation}
relative to $(\Delta z')^2$ of all other wave packets parametrized 
by $0<\varepsilon<\infty$.

\noindent The history of such a minimal wave packet most closely 
resembles a particle world line of finite length $t'$. Such a wave
packet has initial size
\begin{equation}
2a=2\sqrt{\frac{t'}{k} }~~,
\label{eq:initial size}
\end{equation}
expands at the (asymptotic) rate
\begin{equation}
2\frac{a}{b}=2\sqrt{\frac{1}{t'k}} ~~(=2\varepsilon)
\label{eq:expansion rate}
\end{equation}
and after proper time $t'$ has full width
\begin{equation}
2\Delta z'_{min}=2\sqrt{\frac{2t'}{k} }~~.
\label{eq:minimum size}
\end{equation}
Here 
\[
k=\sqrt{\frac{m^2c^2}{\hbar^2} +k_y^2 +k_x^2}
\]
is the Compton wave number of the wave packet of a particle 
with non-zero transverse motion. With increasing  particle
mass the expansion rate $\varepsilon =\sqrt{1/t'k}$
tends towards zero, and, with it, the expansion rate and the 
size of the minimal wave packet. In this limit this corresponds
to the razor sharp world line of a classical particle. This 
circumstance is what is summarized by Eq.(\ref{eq:particle limit}).

Metaphysically, however, the rest mass of a particle is always finite.
Thus the size of its wave packet and the concomitant
expansion rate are always non-zero nevertheless. As an example, take
the world line of a neutron in a state of rest for the time duration
$t'_{conv}=1$ second in conventional units. The neutron's Compton wave
length is $\lambda_n =2\times 10^{-14}$ cm. Consequently, its minimum
initial wave packet size is
\[
2a=2\sqrt{\lambda_n c t'_{conv}}=4.6\times 10^{-2}~~\textrm{cm}~~.
\]
After one second the full width of this wave packet is
\[
2\Delta z_{min}\equiv 2.8 a = 6.4\times 10^{-2} ~~\textrm{cm}~~,
\]
and it expands at the nonzero rate
\[
v_{conv} \equiv 2\frac{a}{b}c=2\sqrt{\frac{\lambda_n c}{t'_{conv}}}
=5\times 10^{-2}~~\textrm{cm/sec}~~,
\]
which is nonrelativistic.

The establishment, and hence the definition, of an inertial frame is done
in terms of Newton's first law of motion.
The classically precise definition of an inertial frame is as follows
\cite{Taylor and Wheeler}:

\noindent \emph{A reference frame is said to be an ``inertial'' or 
  ``free-float'' or ``Lorentz'' reference frame in a certain region of
  space and time when, throughout that region of spacetime -- \emph{and
  within some specified accuracy}-- every free test particle initially
  at rest initially at rest with respect to that frame remains at
  rest, and every test particle initially in motion with respect to
  that frame continues its motion without change in speed or in
  direction.}

The fact that every test particle has a finite Compton radius and
hence is described by a moving wave packet of non-zero minimal width
and expansion rate, implies that an inertial frame cannot be made
arbitrarily small.  Nevertheless, we adopt the above definition and
extend it to quantum mechanics by replacing the (mathematically
uncountable) set of test particles with the (countable) complete set
of Klein-Gordon-orthonormal wave packet histories
\begin{equation}
\{ \phi^{\varepsilon \pm}_{j\ell}(kU,kV)~:~j,\ell= 0,\pm 1,\pm 2, \cdots \},
\end{equation}
which are governed by the relativistic wave equation (\ref{eq:1}).

Let us use these wave packet histories to establish an inertial frame.
We do this by replacing the parallel particle world lines
used by Marzke and Wheeler\cite{Marzke and Wheeler} with corresponding
wave packets.

Without specifying the construction of a quantum mechanical clock, we
consider for the duration of a finite time interval, say $t'$, an
array of $N$ freely floating material particles (resp. anti-particles)
which
\begin{enumerate}
\item[(i)]
have the same proper mass
\item[(ii)]
have no velocity relative to one another
\item[(iii)]
have a minimal separation which is large enough to allow them 
to be distinguished by their spatial location.
\end{enumerate}
This separation is the spacing between the contiguous wave
packet histories 
\[
\phi^{\varepsilon \pm}_{1\ell}(kU,kV),\phi^{\varepsilon \pm}_{2\ell}(kU,kV),
\cdots ,\phi^{\varepsilon \pm}_{N\ell}(kU,kV)
\]
for each of these particles (resp. anti-particles).
Equation (\ref{eq:proper halfwidth}) tells us that this separation is
\[
\frac{2\pi}{k\varepsilon}=\frac{2\pi}{\varepsilon}
\frac{mc}{\hbar}~~.
\]
Optimized relative to the time interval $t'$, these wave packets
expand at the rate $\varepsilon =\sqrt{1/t'k}$. As a result,
their optimal (i.e. minimum) separation is
\begin{equation}
\frac{2\pi}{k\varepsilon}=2\pi \sqrt{\frac{t'}{k}}~~~(=2\pi a )
\label{eq:minimum separation}
\end{equation}
The total proper spatial extent of these $N$ freely floating particles
is therefore
\begin{eqnarray}
2\pi \sqrt{\frac{t'}{k}}N&=&2\pi \sqrt{\frac{t'\hbar}{mc}}N\\
                         &\equiv&L~~.
\end{eqnarray}
They make up a meter rod of length $L$, whose resolution and accuracy
cannot be improved upon except by increasing the mass of its
constitutive particles or by decreasing the amount of time that this
meter rod is used. More precisely, although the \emph{absolute}
accuracy, Eq.(\ref{eq:minimum separation}), used to define an inertial
frame is limited by these two quantities, the \emph{relative} accuracy
can be increased by making $N$, and hence the spatial extent $L$,
sufficiently large.

\subsubsection{Rindler Frame Limit}

\noindent
The behaviour of a nonrelativistically expanding wave packet is
interesting.  One can make it expand faster by having its minimum
proper size, Eq.(\ref{eq:proper width}) be smaller (or having its
transverse wave number $k$ be smaller). Suppose one keeps increasing
$\varepsilon$ until the expansion rate, Eq.(\ref{eq:proper expansion
  rate}), becomes highly relativistic, i.e. $1\ll \varepsilon$.  Under
this circumstance the inertial-frame-limit analysis would no longer
apply. Instead, a glance at Eq.(\ref{eq:linear map omega}) tells us
that the wave packet $\phi^{\varepsilon \pm}_{j\ell}(kU,kV)$ starts
resembling a Minkowski-Bessel mode, and the evolution process of the
wave packet starts resembling the Lorentzian Mach-Zehnder interference
process, Figure \ref{fig:two interferometers}, whose details are
depicted in Figures \ref{fig:amplitude splitting},\ref{fig:reflected
  modes}, and \ref{fig:WKB modes}.
  
However, this resemblance is a peculiar one: The M-B mode oscillates
infinitely rapidly at and near $U=V=0$, the intersection of the two
horizons, while the wave packet $\phi^{\varepsilon
  \pm}_{j\ell}(kU,kV)$ is smooth at this event, even when $1\ll
\varepsilon$. This persistent discrepancy implies that \emph{each wave
  packet \emph{bifurcates} into a pair of wave packets, one in Rindler
  Sector $I$, the other in Rindler Sector $II$, which then
  \emph{re-coalesce} into a single wave packet in $F$}. This
bifurcation-recoalescence process is an interference effect. It is
exhibited by each relativistically collapsing and re-exploding wave
packet $\phi^{\varepsilon \pm}_{j\ell}(kU,kV)$.  Its squared modulus ,
as calculated in the tutorial of Eq.(\ref{eq:squared modulus}), is
controlled by the (squared) \emph{constructive interference parameter}
\begin{equation}
\delta^2\equiv
\Delta^2 \left(\frac{\partial S}{\partial \overline{\omega}} \right)^2
  \left[
     1+ \left(
             \Delta^2 \frac{\partial^2 S}{\partial \overline{\omega}^2}
        \right)^2
  \right]^{-1}
~~~~~~~~~~~~~~~~~
\left(
  \Delta \equiv \frac{2\pi}{\varepsilon} \ll 1 
\right)
\label{eq:constructive interference parameter}
\end{equation}
(\emph{Notation:} $\partial /\partial \overline{\omega}$ means first
take the partial with respect to $\omega$ and then evaluate at it at
$\overline{\omega}$.) The region of spacetime where the wave packet has
non-negligible amplitude is characterized by the in equality
\begin{equation}
-1\le \delta \le 1 ~~.
\label{eq:delta interval}
\end{equation}
With the help of Figure \ref{fig:WKB modes} one finds that
\begin{equation}
\Delta^2 \frac{\partial^2 S}{\partial \overline{\omega}^2}
=\pm \frac{\Delta^2}{\sqrt{\overline{\omega}^2 +k^2 UV}}
\end{equation}
This term, or rather its square, is negligibly small in
Eq.(\ref{eq:constructive interference parameter}) throughout the
spacetime region of interest, namely wherever the oscillatory WKB approximation is
applicable (away from the hyperbolic boundaries between the
classically allowed and the classically forbidden regions in Rindler
Sectors $I$ and $II$).  With this approximation, together with Figure
\ref{fig:WKB modes} and Eqs.(\ref{eq:retarded})-(\ref{eq:advanced})
one finds that Eq.(\ref{eq:constructive interference parameter}) is
equivalent to one of the straight lines
\begin{equation}
(t-t_0)\sinh\left(\ell \varepsilon +\delta\frac{\varepsilon}{2\pi}\right)+
(z-z_0)\cosh\left(\ell \varepsilon +\delta\frac{\varepsilon}{2\pi}\right)=
\frac{\overline\omega}{k}\frac{V}{\vert V\vert}
;~~~~~~\overline{\omega}=\frac{2\pi}{\varepsilon}j~~~~j=0,\pm 1,\pm 2,
\cdots
\label{eq:V straight line}
\end{equation}
when 
\begin{eqnarray*}
S&\equiv&{\mathcal{S}}^V -\omega \ell \varepsilon\\
&=&\omega \ln(\omega+\sqrt{\omega^2+k^2UV})-
                \sqrt{\omega^2+k^2UV}-\omega\ln k\vert V \vert -\frac{\pi}{4}
-\omega \ell \varepsilon~~,
\end{eqnarray*}
 or to 
\begin{equation}
(t-t_0)\sinh\left(\ell \varepsilon +\delta\frac{\varepsilon}{2\pi}\right)+
(z-z_0)\cosh\left(\ell \varepsilon +\delta\frac{\varepsilon}{2\pi}\right)=-
\frac{\overline\omega}{k}\frac{U}{\vert U\vert}~~~~~~~~~~~~~~~~~~~~~~~~~~~~~~~~~~~~~~~
\label{eq:U straight line}
\end{equation}
when 
\begin{eqnarray*}
S&\equiv&{\mathcal{S}}^U -\omega \ell \varepsilon\\
&=&\omega \ln(\omega+\sqrt{\omega^2+k^2UV})-
                \sqrt{\omega^2+k^2UV}-\omega\ln k\vert U \vert -\frac{\pi}{4}
-\omega \ell \varepsilon~~.
\end{eqnarray*}
These straight lines form a $\delta$-parametrized family whose
parameter is restricted by Eq.(\ref{eq:delta interval}). As we know,
this restriction guarantees that the events on these straight lines lie in those
regions of spacetime where the wave packet $\phi^{\varepsilon
  \pm}_{j\ell}(kU,kV)$ has non-negligible amplitude. As the interference parameter
varies from $\delta=-1$ to $\delta=1$, the family of straight lines
``generates'' the spacetime domain of constructive interference. Outside
this domain the wave packet has exponentially small amplitude.

With the help of Eq.(\ref{eq:Rindler I}) the family of generating
lines is represented in Rindler Sector $I$ by
\[
\xi \cosh \left(\tau +\ell \varepsilon +\delta\frac{\varepsilon}{2\pi} \right)=\frac{\vert \overline{\omega}\vert}{k}~~,
\quad \vert \delta \vert \le 1~~.
\]
Consequently, the region swept out by this $\delta$-parametrized family consists of the
set of events whose intersection with Rindler Sector $I$ equals
\[
\{(\xi ,\tau):~\xi \cosh \left(\tau +\ell\varepsilon +\delta\frac{\varepsilon}{2\pi}\right)=\frac{\vert \overline{\omega}\vert}{k};~~
\vert \delta \vert \le 1 \}~~.
\]
This set is obviously bounded away from the event $U=V=0$. In fact,
its proper distance away from this event is 
\[
\frac{\overline{\omega}}{k }\frac{1}{\cosh\varepsilon/2\pi} ~~.
\]
In Rindler Sector $II$ there exists an inverted image ($U \to -U,~V\to
-V$) of this set.  The proper separation between these two sets is
therefore 
\begin{equation}
\left(
\begin{array}{c}
\textrm{separation~between}\\
\textrm{bifurcated~wave}\\
\textrm{packet~histories}
\end{array}
\right)
= 2\frac{\overline{\omega}}{k }\frac{1}{\cosh\varepsilon/2\pi} ~~.
\label{eq:bifurcation distance}
\end{equation}
The spacetime between these two sets is an ``island of calmness''
(zero wave packet amplitude) whose center is the event $U=V=0$ where
every wave packet has strictly zero amplitude.

\subsection{Collapsing Wave Packet Bifurcates: Double Slit}

It is qualitatively obvious that a relativistically collapsing wave
packet should bifurcate into a pair of distinct wave packets, which
upon re-exploding merge into a single relativistically expanding wave
complex. The reason is destructive interference between the
Minkowski-Bessel modes comprising the wave packet. They lie in a
narrow range of Rindler frequencies and, as demonstrated in one of the
ensuing tutorials (Section IX), their amplitudes add or cancel depending
on the location in spacetime.

Recall that every M-B mode oscillates with infinite rapidity as one
approaches the intersection $(U=V=0)$ of the future and past event
horizons. This is evident from Figures \ref{fig:amplitude splitting}
and \ref{fig:WKB modes}. When superimposed, such rapidly oscillating
amplitudes interfere destructively whenever that superposition extends
over a small but finite range of Rindler frequencies. Moreover, that
destructive interference becomes perfect as one approaches the event
$U=V=0$ from any direction in spacetime.  This event is the center of
an island of ``total calmness'' (zero amplitude) in spacetime.  This
island, whose spacelike extent is given by Eq.(\ref{eq:bifurcation distance}),
separates the pair of wave packet histories, one in Rindler Sector
$I$, the other in Sector $II$, before they coalesce into the single
history of a relativistically exploding wave packet in Rindler Sector
$F$.The boundary of
this packet expands with asymptotic velocity whose magnitude (relative
to the center of this packet) is
\[
u=\tanh \frac{\varepsilon}{2\pi}~~,
\]
a fact which follows from Eq.(\ref{eq:V straight line}) or 
(\ref{eq:U straight line}) by letting $\ell=0$.

On each spacelike slice of elapsed time ($\xi=const.$) in $F$ through
this history, this wave complex consists of a pattern of interference.
This interference is between the two amplitudes of the bifurcated wave
packet amplitudes from sectors $I$ and $II$ respectively. Thus, in a
physically rigorous sense, the two Rindler sectors $I$ and $II$ act as
a double slit arrangement which accommodates an interference process
between coherent amplitudes passing through two finite spacelike
openings separated by the above-mentioned ``island of calmness''.

Both ``double slit'' and ``Lorentzian Mach-Zehnder interferometer''
are accurate terms which describe the spacetime arrangement that gives
rise to the interference process $P\rightarrow (I,II)\rightarrow F$.
The difference is that the latter term highlights the reflection
process which is brought about in $I$ and $II$ by their respective
pseudo gravitational potentials. The former term leaves the nature of
this reflection process unspecified.

\subsection{Accelerons}

We know that, when $\varepsilon\gg 1$, the wavelet
$\hat\phi_{j\ell}^\varepsilon(\omega)$ is highly localized around a
well-defined Rindler frequency, say $\omega=\overline{\omega}$. The
concomitant wave packet history is
\begin{equation}
\phi^{\varepsilon \pm}_{j\ell}(kU,kV)
        =\int_{-\infty}^\infty 
                \left[ \hat \phi^\varepsilon_{j\ell} (\omega)\right]^*
        B^\pm_\omega(kU,kV) ~d\omega~~~.
\end{equation}
    Consequently, the wave amplitude of $\phi^{\varepsilon 
    \pm}_{j\ell}(kU,kV)$ and that of $B^\pm_{\overline\omega}(kU,kV)$
    are roughly the same functions of spacetime $(U,V)$. A glance at
    Figures \ref{fig:amplitude splitting},\ref{fig:reflected modes},
    or \ref{fig:WKB modes} reveals therefore that
    $\phi^{\varepsilon\pm}_{j\ell}(kU,kV)$ is non-zero in $F$, in $P$, as well
    as in those regions of Rindler sectors $I$ and $II$ which lie
    between the two conjugate hyperbolas (``boundaries of
    evanescence'') and the event horizons.

These mathematical properties of this wave amplitude express the following
physical meaning:

\noindent
First of all, relative to any globally inertial reference frame
$\phi^{\varepsilon \pm}_{j\ell}(kU,kV)$ expresses the history of a wave
complex which is \emph{finite}. The two timelike hyperbolic boundaries
of evanescence, indicated in Figures \ref{fig:amplitude
splitting},\ref{fig:reflected modes}, and \ref{fig:WKB modes} guarantee
this.

\noindent
Second, this wave complex collapses \emph{relativistically to a
minimum size}, the proper distance between the two hyperbolic
boundaries, before it \emph{explodes relativistically}. Third,
relative to accelerated observers in Rindler Sectors $I$ and/or $II$,
the wave complex is an approximately \emph{static} entity. Its Rindler
lifetime is very long:
\[
\Delta\tau\approx \varepsilon \gg1
\]
During this time interval the boundary, and hence the size
\begin{equation}
\xi=\frac{2\overline{\omega}}{k};~~~~~
\overline{\omega}=\frac{2\pi}{\varepsilon}j~~~~j=0,\pm 1,\pm 2,
\cdots
\label{eq:slabthickness}
\end{equation}
of this entity is approximately static. In this article the boundary
consists of two parallel planes parallel to the $y$ and $x$
coordinates; in other words, we have a static slab \cite{Gerlach1996} whose
proper thickness is Eq.(\ref{eq:slabthickness}). In its
\emph{interior} the wave field oscillates with mean Rindler frequency
$\overline{\omega}$.

Thus $\phi^{\varepsilon \pm}_{j\ell}(kU,kV)$ is a field configuration which
\begin{itemize}
\item vibrates
\item is localized along the $z$-direction, and
\item has a static boundary
\end{itemize}

What is the field amplitude (or the quantum states) of this
oscillating entity?  This is a question one would ask relative to the
two accelerated frames, Rindler $I$ and $II$, and it pertains to the
field amplitudes and phases of these two sectors.  The answer is
mathematically simple 
\cite{1}, 
but it needs to be justified
also physically, i.e. related to observations accessible to an
inertial observer. \emph{If one can do this}, then the above oscillating
entity has a fourth property, namely $\phi^{\varepsilon \pm}_{j\ell}(kU,kV)$
\begin{itemize}
\item is a Klein-Gordon degree of freedom with quantum states of excitation.
\end{itemize}

We shall call a wave complex with the above four properties an \emph{acceleron}.
This is because of the accelerative nature of its boundary relative to
an inertial frame.

In summary, there are two kinds of wave mechanical building blocks in
nature.  Wave packets ($\varepsilon \ll 1$) and accelerons
($\varepsilon \gg 1$).  Wave packets express the particle-like
properties of ponderable matter.  (The deflection of light by the sun
is an example.) This is because these properties are measured by an
observing apparatus based on the intrinsic attributes of a free float
frame. By contrast, accelerons express those ``complementary''
properties of ponderable matter (Amplification of light passing through
on accelerated dielectric, see Section X.B), which are measured by an observing apparatus based on
the attributes of a pair of frames accelerating uniformly and linearly
into opposite directions.

The wave packet properties and the acceleron properties are
``complementary'' in the sense of wave mechanics. Indeed, wave packets
have phase space area representatives which are tall and skinny
($\varepsilon \ll 1$), while accelerons have phase space
representatives which are short and squatty ($\varepsilon \gg 1$),
as in Figure \ref{fig:cells_skinny_and_squatty}.
The vertical and the horizontal phase space coordinates quantify
properties (Minkowski Frequency and Rindler Frequency) which are
experimentally mutually exclusive. Their dichotomic nature is
captured by the phrase ``complementary''.

\section{MATHEMATICAL TUTORIAL: PHASE SPACE OF SQUARE-INTEGRABLE FUNCTIONS}

\subsection{Why}

The motivation for introducing the phase space of square-integrable
functions is both physical and mathematical.

The mathematical motivation comes from the necessity that we
comprehend the square-integrable test functions and their Fourier
transforms from a \emph{single} point of view. This means that the
classification of these functions be based on their behavior on the
given domain concurrent with the behaviour of their transform on the
Fourier domain.

The physical motivation comes from the necessity that we identify the
Klein-Gordon dynamics in terms of orthonormal degrees of freedom which
are localized both in their spacetime-translation-induced properties
(momentum) and in their Lorentz-boost-induced properties (boost
``energy''). The momentum and the boost ``energy'' domains are Fourier
complementary attributes. One indicates an inertial frame, the other a
pair of frames accelerating into opposite directions. Consequently,
these complementary attributes seem to constitute a metaphysically
insurmountable either-or duality.
Nevertheless, comprehending nature demands that we represent any given
Klein-Gordon solution in terms of orthonormalized degrees of freedom
which capture \emph{both} attributes in the manner dictated by the Fourier
transform.

The phase space of functions square-integrable on the real line not only
fulfills this demand but also answers the following two questions:
\begin{enumerate}
\item How does one construct Klein-Gordon orthonormal wave histories?
\item How does one obtain a physical classification of the solutions
to the relativistic wave equation?
\end{enumerate}
The ``physical'' classification is to include not only the
circumstance of low dispersion wave packets tracing out a narrow and
straight world tube, but also the circumstance of wave packets with
such high dispersion that they explode relativistically and thus have
a non-trivial internal dynamics.

\subsection{Phase Space Construction}

Instead of developing that phase space in its most general
mathematical form, we shall illustrate its basic nature with an
archetypical construction. This is a three step process:
\begin{enumerate}
\item[(1)]
First, exhibit a complete set of orthonormal wave packet functions and their
Fourier transforms.
\end{enumerate}

This first step is facilitated by noting that there is a practical and
general method for constructing square-integrable functions which are
orthonormal on the real line $-\infty <\theta <\infty$. To use this
method, one needs to pick only a single square-integrable function
with an easily specifiable property stipulated by the following
theorem:

\begin{em}
For any square-integrable function $\phi(\theta)$ the set $\{ \phi_\ell
(\theta) \equiv \phi (\theta -\frac{2\pi}{\varepsilon}\ell):~~
\ell=0,\pm 1,\pm 2, \cdots\}$
is an orthonormal system, i.e.
\[
(\phi_\ell,\phi_{\ell'})\equiv \int_{-\infty}^\infty
\phi_\ell (\theta)^*\phi_{\ell'}(\theta)~d\theta=\delta_{\ell {\ell'}}~~,
\]
if and only if the Fourier transform of $\phi(\theta)$,
\[
\hat \phi(\omega)=\frac{1}{ \sqrt{2\pi} }\int_{-\infty}^\infty \phi(\theta)
e^{i\omega\theta}~~d\theta
\]
satisfies 
\[
\sum_{n=-\infty}^\infty \vert \hat \phi(\omega +\frac{2\pi}{\varepsilon}n)\vert^2=
\frac{\varepsilon}{2\pi}~~.
\]
\end{em}
\noindent
This powerful theorem, well known to workers in wavelet theory
\cite{Louis}, captures the key idea behind the success in using
windowed Fourier transforms (among others) to construct orthonormal
wavelets. Indeed, let us apply this theorem to the windowed function
\[
\phi(\theta)= \frac{1}{\sqrt{\varepsilon}} 
\exp \left(-i\frac{2\pi}{\varepsilon}j\theta \right)\times \left\{
\begin{array}{ccc}
1& ~~~~& -\frac{\varepsilon}{2} \le \theta \le \frac{\varepsilon}{2}\\
0& ~~~~& 0<\frac{\varepsilon}{2}<\vert \theta \vert
\end{array}
\right.
\]
Here $j$ is an arbitrary integer.

\noindent
The Fourier transform of this function is
\[
\hat \phi (\omega)=   \frac{1}{\sqrt {2\pi\epsilon}}
   \frac{2 \sin \left(  \omega - {2\pi \over \epsilon}j \right) 
          \frac{\epsilon}{2} }{\left(  \omega - {2\pi \over \epsilon}j\right)}
\]
and the sum mentioned in the theorem is
\begin{eqnarray*}
\sum_{n=-\infty}^\infty \vert \hat \phi(\omega +\frac{2\pi}{\varepsilon}n)\vert^2
&=&
\frac{1}{2\pi \epsilon} \sum_{n=-\infty}^\infty 
\frac{4 \sin^2 \left(  
                \omega - {2\pi \over \epsilon}j +{2\pi \over \epsilon}n \right) 
          \frac{\epsilon}{2} }{\left(  
        \omega - {2\pi \over \epsilon}j +{2\pi \over \epsilon}n\right)^2}\\
&=&\frac{\varepsilon}{2\pi} \sin ^2 \frac{\omega \varepsilon}{2}
\sum_{n=-\infty}^\infty \frac{1}{\left( \frac{\omega \varepsilon}{2}+\pi n\right)^2}\\
&=&\frac{\varepsilon}{2\pi}~~\textrm{for~all}~~-\infty<\omega <\infty .
\end{eqnarray*}
This means we can use the theorem. It guarantees us that the system 
of exponentials windowed
along the $\theta$-axis,
\be
\phi_{j\ell}(\theta)= \frac{1}{\sqrt{\varepsilon}} 
\exp \left(-i\frac{2\pi}{\varepsilon}j\theta \right)\times \left\{
\begin{array}{ccc}
1& ~~\textrm{whenever}& (\ell-\frac{1}{2})\varepsilon \le \theta \le (\ell+\frac{1}{2})\varepsilon\\
0& ~~\textrm{whenever}& \frac{\varepsilon}{2}<\vert \ell \varepsilon 
-\theta \vert
\end{array}
\right\} ~~~j,\ell=0,\pm1,\pm,2,\cdots~~,
\label{eq:windowed exponential}
\ee
and their Fourier transforms
\begin{eqnarray}
\hat \phi_{j\ell} (\omega)
&=&
\frac{1}{\sqrt {2\pi \varepsilon}}
\int _{(l-{1\over 2})\epsilon }^{(l+{1\over 2}) \varepsilon } 
             \exp {i\left( \omega - {2\pi \over \varepsilon}j\right) \theta }
     ~d\theta 
\quad ~~~~~~~~~~~~~~~~~~~~~~~~~~~~~~~j,\ell=0,\pm 1,\cdots \nonumber\\
&=&
\frac{1}{\sqrt {2\pi \varepsilon}}
\exp {i\left(  \omega- {2\pi \over \varepsilon}j \right) l\epsilon } \times
   \frac{2 \sin \left(  \omega- {2\pi \over \epsilon}j\right) 
          \frac{\epsilon}{2} }{\left(  \omega-{2\pi \over \epsilon}j \right)},
\label{eq:Fourier windowed exponential}
\end{eqnarray}
form an $\varepsilon$-parametrized family of sets of wave packets orthonormal on
the Fourier-complementary domains $-\infty<\theta<\infty$ and
$-\infty<\omega<\infty$:
\begin {equation}
\int_{-\infty}^\infty \phi_{j\ell}(\theta)^*\phi_{j'\ell'}(\theta)~d\theta=
\int_{-\infty}^\infty \hat \phi_{j\ell}(\omega)^*\hat \phi_{j'\ell'}(\omega)~d\omega=
\delta_{jj'}\delta_{ll'}~~~.
\end{equation}

\begin{enumerate}
\item[(2)]
Second, assemble the function domain and the corresponding Fourier domain into
a two-dimensional phase plane, and use the localized nature of these
functions to partition this plane into a set of mutually exclusive and
jointly exhaustive phase space cells.
\end{enumerate}

This step consists of taking note of the localized nature of the orthonormal
basis functions and their Fourier transforms. From their definitions one finds
that
\[
\phi_{j\ell}(\theta) ~\textrm{is~concentrated~in~}(\ell-\frac{1}{2})\varepsilon <\theta<
(\ell+\frac{1}{2})\varepsilon,
\]
while its Fourier transform
\[
\hat \phi_{j\ell}(\omega) ~\textrm{is~concentrated~in~}
(j-\frac{1}{2})\frac{2\pi}{\varepsilon} <\omega<(j+\frac{1}{2})\frac{2\pi}{\varepsilon}.
\]
It follows that the phase space of square-integrable functions
\begin{enumerate}
\item[(i)]
consists of the two-dimensional plane spanned by the cartesian coordinates
$-\infty<\theta<\infty$ and $-\infty<\omega<\infty$ and
\item[(ii)]
is partitioned into elements of area
\[
\varepsilon \times \frac{2\pi}{\varepsilon}=2\pi
\]
whose horizontal width $\varepsilon$ and vertical height $\frac{2\pi}{\varepsilon}$
are the separations between the adjacent wave packets $\phi_{j\ell}(\theta)$
and $\hat \phi_{j\ell}(\omega)$. The centers of these area elements constitute
the lattice work of points
\[
\{~(\theta ,\omega)=\left(\ell \varepsilon, j \frac{2\pi}{\varepsilon} \right):~~
\ell ,j=0,\pm1,\pm,2,\cdots~\}
\]
where the modulus of the wavelets $\phi_{j\ell}(\theta)$ and $\hat \phi_{j\ell}(\omega)$
have their respective maximum values.
\end{enumerate}
Roughly speaking, the o.n. wavelets $\phi_{j\ell}(\theta)$ and their
Fourier transforms $\hat \phi_{j\ell}(\omega)$ cause the phase space
to be covered completely with a set of rectangular tiles of equal area
whose shape is determined by the parameter $0<\varepsilon<\infty$.
This is depicted in Figure~\ref{fig:cartesian phasespace}

\begin{enumerate}
\item[(3)]
Third, use the completeness and the orthonormality properties to assign to these cells
the respective Fourier coefficients of any given square-integrable function.
\end{enumerate}
This final step consists of taking note of the completeness
of each set of o.n. wavelets. Thus any square-integrable function
$f(\theta)$ and its Fourier transform $\hat f(\omega)$ is a linear combination
of these wavelets
\begin{eqnarray}
f(\theta)&=& 
\sum_{j=-\infty}^\infty \sum_{\ell=-\infty}^\infty
\phi_{j\ell}(\theta) \int_{-\infty}^\infty 
           \phi_{j\ell}^*(\theta')f(\theta')~~d\theta' \\
\hat f(\omega)&=& 
\sum_{j=-\infty}^\infty \sum_{\ell=-\infty}^\infty
\hat \phi_{j\ell}(\omega) \int_{-\infty}^\infty 
           \hat \phi_{j\ell}^*(\omega')\hat f(\omega')~~d\omega'
\end{eqnarray}
It is remarkable and obvious (because they are Fourier transforms of
each other) that even though these two functions are superficially
different, their Fourier coefficients are the \emph{same}:
\begin{eqnarray}
\int_{-\infty}^\infty 
           \phi_{j\ell}^*(\theta')f(\theta')~~d\theta'&=&
\int_{-\infty}^\infty 
           \hat \phi_{j\ell}^*(\omega')\hat f(\omega')~~d\omega'\\
&\equiv& c_{j\ell}
\end{eqnarray}
Consequently, the Fourier transform pair has the form
\be
f(\theta)=
\sum_{j=-\infty}^\infty \sum_{\ell=-\infty}^\infty c_{j\ell} \phi_{j\ell}(\theta) 
\ee
and
\be
\hat f(\omega)=
\sum_{j=-\infty}^\infty \sum_{\ell=-\infty}^\infty c_{j\ell} \hat \phi_{j\ell}(\omega)~. 
\ee
These two representations imply that any square-integrable function,
together with its Fourier transform, is represented geometrically
by assigning each common Fourier coefficient $c_{j\ell}$ to its
phase space area element located at $(\theta,\omega)=(\ell\varepsilon,
j\frac{2\pi}{\varepsilon})$. Even though the magnitude of the area of
all these elements is the same, namely $2\pi$, their shape depends on the
parameter $\varepsilon$. As shown in 
Figure~\ref{fig:cells_skinny_and_squatty}, when $\varepsilon\ll1$, they
are compressed along the horizontal and stretched along the vertical,
and vice versa when $1\ll\varepsilon$.

\subsection{Exploding Wave Packets: Relativistic Internal Dynamics ($\varepsilon \gg 1$)}

When $1\ll\varepsilon $ then each phase space elements gets mapped 
by Eq.(\ref{eq:linear map omega}) into
a relativistically collapsing and re-exploding wave complex
whose spacetime support is indicative of the partitioning of spacetime
into the four Rindler sectors, Figure \ref{fig:Rindler spacetime}. The
collapse and the re-explosion is so violent as to preclude a wave
mechanical description of a classical particle relative to an inertial
frame. Instead, the physical (i.e. measurable) properties of the
evolution process must be described in terms of the Lorentz version of
the Mach-Zehnder interference process of Section $I$. For every pair
of integers $(j,\ell)$ the Klein-Gordon solution $\phi^{\varepsilon
\pm}_{j\ell}(kU,kV)$ expresses such a process.

The wave complex $\phi^{\varepsilon
\pm}_{j\ell}(kU,kV)$ is the relativistic antithesis of any one of the
familiar plane wave packets. Each complex
\begin{enumerate}
\item
has an indeterminate direction in the $z-t$ plane but has a
well-defined mean Rindler (``boost'') frequency 
\[
{\overline \omega}= \frac{2\pi}{\varepsilon}j~~,\quad j=0,\pm 1,\pm 2,\cdots ~~~,
\]
\item
has a Rindler time life span of order
\[
\Delta\tau\approx \varepsilon \gg1~~,
\]
which is centered around
\[
\tau=\ell \varepsilon~~, \quad \ell=0,\pm1,\cdots ~~,
\]
\item has an ill-defined group velocity which is less sharp the larger
  $\varepsilon$ is, and
\item
has a very large asymptotic velocity spread whose magnitude in the wave 
packet rest frame is
\be
\Delta u= 2 \tanh \frac{\varepsilon}{2\pi}\quad 
                                        (\textrm{``relativistic~velocity~spread''})~~.
\ee
\end{enumerate}

\section{MATHEMATICAL TUTORIAL: WAVE PACKETS VIA CONSTRUCTIVE INTERFERENCE}

Let us describe and apply a general method which facilitates the
extraction of the physical properties of wave packets histories from
their integral representations, Eqs.(\ref{eq:linear map omega}) and
(\ref{eq:linear map theta}). Their explicit form is
\begin{eqnarray}
\phi^{\varepsilon \pm}_{j\ell}(kU,kV)
        &\equiv&\int_{-\infty}^\infty   
                \left[
\frac{1}{\sqrt {2\pi \varepsilon}}
\exp {i\left(  \omega- {2\pi \over \varepsilon}j \right) l\epsilon } \times
   \frac{2 \sin \left(  \omega- {2\pi \over \epsilon}j\right) 
          \frac{\epsilon}{2} }{\left(  \omega-{2\pi \over \epsilon}j \right)}
                \right]^*
        B^\pm_\omega(kU,kV) ~d\omega  \label{eq:exploding packet}\\
        &=&\int_{(\ell-\frac{1}{2})\varepsilon}^{ (\ell+\frac{1}{2})
          \varepsilon}
                        \left[
        \frac{1}{\sqrt{\varepsilon}} \exp \left(-i\frac{2\pi}{\varepsilon}j\theta \right)
                        \right]^*
\frac{1}{ \sqrt{2\pi} }e^{\mp ik(Ve^\theta +Ue^{-\theta})/2}  d\theta
~~~j,\ell =0,\pm1,\pm2,\cdots~~~, \label{eq:expanding packet}
\end{eqnarray}
These are different representations of one and the same wave packet
$\phi^{\varepsilon \pm}_{j\ell}(kU,kV)$. Note that each representation
has its own window function. In Eq.(\ref{eq:exploding packet}) there
is the ``sinc'' function whose half width in the $\omega$-domain is
$2\pi/\varepsilon$. By contrast in Eq.(\ref{eq:expanding packet})
there is the ``square wave pulse'' function whose width in the
$\theta$-domain is $\varepsilon$. Consequently, for large $\varepsilon$
the representation (\ref{eq:exploding packet}) has a
narrow window and the representation (\ref{eq:expanding packet}) a
wide window, while for small $\varepsilon$ it is the
other way around.

\subsection{Overview}
The method guarantees
success for both large-$\varepsilon$ (relativistically collapsing and
exploding) and for small-$\varepsilon$ (slowly contracting and expanding)
wave packets. The basic philosophy underlying this method
is akin to the method of steepest descent, and it is the same for
Eq.(\ref{eq:exploding packet}) and for Eq.(\ref{eq:expanding packet})
In
both cases the given integral is replaced to an excellent
approximation by a Gaussian integral. The method employed here is a
three step procedure.
\begin{enumerate}
\item
Replace the phase modulated window functions (in Eq.(\ref{eq:exploding packet})
or in Eq.(\ref{eq:expanding packet}), whichever one has the narrower maximum) with
its equivalent phase modulated Gaussian window function.
\item
With the location of the maximum as the reference point, expand the
phase of the modes in a Taylor series. The narrowness of the Gaussian
window function guarantees that terms which are cubic or higher may be
omitted without affecting the accuracy of the integral.
\item The outcome of steps 1 and 2 is a Gaussian integral. Evaluate it
  and thereby obtain a slowly varying amplitude times a rapidly
  varying phase factor. This product is the sought-after history of a
  wave complex.
\end{enumerate}
From this modulated Gaussian extract the key properties of the wave packet
history:
\begin{enumerate}
\item
The amplitude locates those events in spacetime where wave field 
quanta can be found. 
\item
By contrast, the rapidly varying phase factor
controls the interference with other wave packets.
\end{enumerate}

\subsection{Evaluation of the Integral}

We shall forego applying this method to the plane waves because the properties
of the resulting slowly expanding ($\varepsilon \ll 1$) wave packets are generally
known. Instead, let us apply it to the relativistically exploding 
($1\ll\varepsilon $) wave packets, Eq.(\ref{eq:exploding packet}).

We replace its Rindler frequency sinc envelope in Eq.(\ref{eq:exploding packet}), 
\[
\frac{ \sin \left( {2\pi \over \epsilon}j -\omega \right) 
          \frac{\epsilon}{2} }{\left( {2\pi \over \epsilon}j -\omega \right)}~,
\]
with the equivalent Gaussian window
\be
\frac{\pi}{\Delta} \exp \left\{ - \frac{( \frac{2\pi}{\varepsilon}j-\omega)^2}
{2 \Delta^2 } \right\} \equiv f(\omega)~~.
\label{eq:Gaussian}
\ee
Here 
\[
\Delta\equiv \frac{2\pi}{\varepsilon}\ll1
\]
is the effective half width of this window.
It has the same maximum amplitude and the same width at its inflection points
as the sinc function.
The corresponding phase modulated Gaussian window function,
\[
\hat \phi_{j\ell}(\omega)\approx \frac{1}{\sqrt{2\pi \varepsilon}}
                        e^{i\left(\frac{2\pi}{\varepsilon}j-\omega \right)\ell
\varepsilon} f(\omega)~~,
\]
takes the place of the exact window function in the history of the
wave complex, Eq.(\ref{eq:exploding packet}).  Its integral
representation is
\begin{eqnarray}
\phi^{\varepsilon\pm}_{j\ell} (kU,kV)
&=&\int_{-\infty}^\infty \left[
        \frac{1}{\sqrt{2\pi \varepsilon}}
        e^{i\left(\frac{2\pi}{\varepsilon}j- \omega \right)\ell \varepsilon} f(\omega)\right]~
{\mathcal{A}}(\omega) e^{i{S}_\omega} ~~d\omega \nonumber\\
&\equiv&
\frac{1}{\sqrt{2\pi \varepsilon}} \int_{-\infty}^\infty {\mathcal{A}}(\omega)f(\omega)e^{iS} ~d\omega
\label{eq:WKB integral}
\end{eqnarray}
where
\begin{equation}
S\equiv \mathcal{S}_\omega -\omega \ell \varepsilon
\label{eq:total phase}
\end{equation}
is the \emph{total} phase and the function $\mathcal{S}_\omega$ is one of the four possible
signed WKB phases, given in Figure \ref{fig:WKB modes}.  The phase
factor $e^{iS}$ is a rapidly varying function of $\omega$.  The slowly
varying function $\mathcal{A}(\omega)$ is the WKB-approximate M-B mode
amplitude (also given in Figure \ref{fig:WKB modes}), which is nearly
constant across the narrow Gaussian $\omega$-window defined by $f(\omega)$,
Eq.(\ref{eq:Gaussian})

Note that the analytic expressions in Figure \ref{fig:WKB modes} apply
only to the ``classically allowed'' region of spacetime, i.e. where
$k^2UV<\omega^2$ between the two conjugate time-like hyperbolas, the
boundaries of the regions of evanescence. We have \emph{not} given, and
we need \emph{not} concern ourselves with those regions of spacetime
where the wave packets have evanescent behaviour. This is because in
those regions the wave packets have exponentially small amplitudes.

The maximum of the Gaussian window, Eq.(\ref{eq:Gaussian}) is located
at the Rindler frequency
\[
\overline{\omega}=\frac{2\pi}{\varepsilon}j \quad \quad j=0,\pm 1,\pm 2, \cdots~~.
\]
The truncated Taylor series of the WKB phase around this point is
\[
S(\omega)=S({\overline{\omega}}) +\frac{\partial S}{\partial
\overline{\omega}} (\omega-\overline{\omega})+\frac{1}{2}
\frac{\partial^2 S}{\partial
\overline{\omega}^2} (\omega-\overline{\omega})^2 ~~.
\]
Introduce this phase into the to-be-evaluated integral, Eq(\ref{eq:WKB integral})
This integral now has the form
\[
\int_{-\infty}^\infty \exp (\alpha z^2 +\beta z)~dz= \sqrt{\frac{\pi}{-\alpha}}
\exp (-\frac{\beta^2}{4\alpha})~~~~~~(Re~\alpha <0)~~.
\]
Use this result and find that the expression for the wave packet is
\begin{equation}
\phi^{\varepsilon\pm}_{j\ell} (kU,kV)= 
\left(
(2\pi^3)^{1/2} \sqrt{\frac{1+i\sigma}{1+\sigma^2} }
\exp\left[ -\frac{\Delta^2}{2}\left(\frac{\partial S}{\partial
\overline{\omega}} \right)^2 \left(\frac{1+i\sigma}{1+\sigma^2}
\right) \right]
\right)
\times 
\left(     \frac{1}{\sqrt{2\pi \varepsilon}}
{\mathcal{A}}(\overline\omega) e^{i{S}({\overline\omega}})
\right)
~~.
\label{eq:final wave packet}
\end{equation}
Here 
\begin{eqnarray*}
\sigma&\equiv& \frac{\partial^2 S}{\partial \overline{\omega}^2}
\Delta^2 \\
&=&\pm \frac{\Delta^2}{\sqrt{\overline{\omega}^2 +k^2 UV}}
\end{eqnarray*}
is negligibly small throughout those regions of spacetime where the
WKB approximation is applicable, i.e. away from the (hyperbolic)
boundaries between the classically allowed and the classically
forbidden regions in Rindler sectors $I$ and $II$.

The utility of this explicit expression for the wave packet history 
is the mathematical transparency of its squared modulus,
\begin{equation}
\vert \phi^{\varepsilon\pm}_{j\ell} (kU,kV) \vert ^2 =
const.\frac{{\mathcal{A}}^2(\overline\omega)}{\sqrt{1+\sigma^2}}
\exp
\left\{ 
  -\Delta^2 \left(\frac{\partial S}{\partial \overline{\omega}} \right)^2 
  \left[
     1+ \left( 
             \Delta^2 \frac{\partial^2 S}{\partial \overline{\omega}^2}
        \right)^2 
  \right]^{-1}
\right\}~~.
\label{eq:squared modulus}
\end{equation}
It answers the mathematical question: In what regions of spacetime
does the wave packet history have non-negligible intensity?
If $\phi^{\varepsilon\pm}_{j\ell} (kU,kV)$ refers to the wave function 
of a single particle (resp. antiparticle), then the squared modulus
answers the physical question: Where in spacetime is there a non-zero
probability for finding a particle (resp. antiparticle)?
The exponent of Eq.(\ref{eq:squared modulus}) gives the necessary
condition. The probability is non-zero in those regions of spacetime where 
the squared \emph{constructive interference parameter}
\begin{equation}
\delta^2\equiv
\Delta^2 \left(\frac{\partial S}{\partial \overline{\omega}} \right)^2
  \left[
     1+ \left(
             \Delta^2 \frac{\partial^2 S}{\partial \overline{\omega}^2}
        \right)^2
  \right]^{-1}
~~~~~~~~~~~~~~~~~
\left(
  \Delta \equiv \frac{2\pi}{\varepsilon} \ll 1 
\right)
\end{equation}
satisfies
\[
\delta^2 \le 1~~~.
\]
This condition is violated in those
regions of spacetime where $\phi^{\varepsilon\pm}_{j\ell} (kU,kV)$ has
negligible amplitude.

\section{DOUBLE SLIT INTERFERENCE \label{sec-INTERFERENCE}}

From a future historical perspective the last quarter of the Twentieth
century will undoubtedly be marked by the intrigue generated by the
acceleration temperature. What other formulas in physics, besides
the acceleration-induced thermal energy
\[
kT=\frac{\hbar}{c}\frac{g}{2\pi}~~,
\]
can rival $E=\hbar\omega$ and $E=c^2m$ in its ubiquity, in its
fundamental role as indicated by the two most important constants of
nature, $\hbar$ and $c$, and in its implication for gravitation
via the surface gravity
\[
g=\frac{c^3}{4MG}
\]
of black holes?

No wonder the interaction between uniformly accelerated quantum
system and the quantized wave field has been the object of such intense
investigations.  The ostensive goal has been to identify the wider
framework which would accommodate our understanding of the acceleration
temperature as given by the Davies-Unruh formula. As a reward, this
framework would identify the iceberg which presumably lies hidden
under the acceleration temperature.

\begin{figure}
\epsfclipon
\epsfxsize=6in
\epsffile[-50 465 570 732]{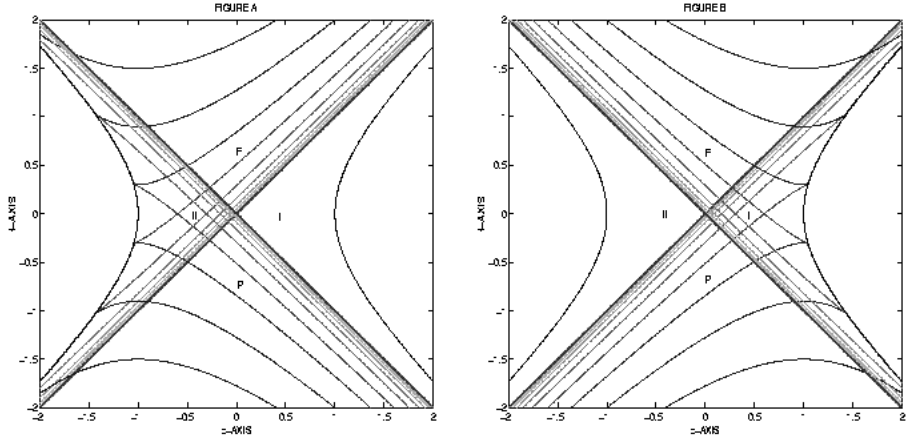}
       \epsfverbosetrue
\caption{Double slit interference of two waves
  modes entering and emerging from a pair of accelerated frames. In
  reality the two figures should be superimposed, one on top of the
  other, in order to show the composite phase fronts of a single
  relativistically collapsing and exploding wave packet.  However, for
  clarity, we show the two interfering waves separately, one passing
  through Rindler Sector $I$ (FIGURE B), the other through Rindler
  Sector $II$ (FIGURE A). The interference occurs in Rindler Sector
  $F$, where the two waves meet and superpose to form a resultant
  wave.  \hfill\break \indent The process of each partial wave passing
  through its respective spacetime sector ($I$ and $II$) is a
  reflection process.  Each figure depicts the WKB phase fronts being
  reflected by the pseudo-gravitational potential. Note that the
  direction of propagation of the wave mode is perpendicular (in the
  Lorentz sense) to its phase fronts. FIGURE B pictures the reflection
  process $P\rightarrow I \rightarrow F$, which is expressed by the
  wave ${\mathcal{A}} \exp [i{\mathcal{S}}^V] +{\mathcal{A}} \exp
  [-i{\mathcal{S}}^U]$, the WKB expression for $B^+_\omega$ in Figure
  \ref{fig:WKB modes}a. The incident wave ${\mathcal{A}} \exp
  [-i{\mathcal{S}}^U]$, which propagates from $P$ towards the boundary
  ($UV=\omega^2$) of evanescence in $I$m represented in the figure by
  the isograms of ${\mathcal{S}}^U$. Similarly, the reflected wave
  ${\mathcal{A}} \exp [i{\mathcal{S}}^V]$, which propagates from the
  boundary of evanescence in $I$ to the future $F$, is represented by
  the isograms of the phase ${\mathcal{S}}^V$. Note that the incident
  and the reflected phase contours intersect the boundary in a
  perpendicular way. \hfill\break \indent In an analogous way, FIGURE A
  pictures the reflection process $P\rightarrow II \rightarrow F$,
  which is expressed by the wave $e^{-\pi \omega}{\mathcal{A}} \exp
  [-i{\mathcal{S}}^U] +e^{-\pi \omega}{\mathcal{A}} \exp
  [i{\mathcal{S}}^V]$, the WKB expression for $B^+_\omega$ in Figure
  \ref{fig:WKB modes}a.
  The superposed process $P\rightarrow (I,II) \rightarrow F$ manifests
  itself as interference in Rindler Sector $F$, where the phase fronts
  in FIGURE A overlap with those in FIGURE B.}
\label{fig:reflected phases}
\end{figure}

\subsection{Accelerated ``Detectors''}

To achieve this goal, the exclusive strategy has been to have the
accelerated quantum system consist of a kind of localized thermometer
which interacts with the thermal ambience in Rindler sector $I$ (or
$II$).

This strategy has been used for the purpose of determining how
an accelerated point particle with internal quantum states responds to a 
relativistic field, 
and, conversely, how the field
responds to the internal dynamics of such a ``detector''.

The simplicity of first order quantum perturbation theory has pretty
much established agreement about the response of an accelerated
``detector'' to the thermal ambience of the Minkowski vacuum.
However, predictions about measurable effects of the ``detector''
dynamics on the field are not characterized by such simplicity.  In
fact, analyses are characterized by higher order perturbation theory
\cite{Unruh and Wald,Massar and Parentani} and/or fairly specialized
tools whose sophistication \cite{Grove,Raine Sciama and
Grove,Unruh,Hinterleitner} is not commensurate with the simplicity of
the phenomenon.

However, an astonishingly simple and fruitful formulation is possible
in terms of interfering wave amplitudes.  Suppose one recognizes that
the distinguishing feature of a Minkowski-Bessel mode propagating from
the past $P$ to the future $F$ is (i) that it splits itself into a
pair of disjoint amplitudes which subsequently recombine coherently in
$F$, and (ii) that this recombination is an interference between the
two amplitudes from the two Rindler Sectors $I$ and $II$. This
(Lorentzian) double slit interference process implies that any quantum
system situated in one or both of these slits (i.e. Rindler sectors)
has a measurable effect on the interference.

\subsection{Amplification by an Accelerated Dielectric}

Instead of considering the familiar single accelerated point particle
with its internal quantum states, take the case of a set of
accelerated oscillators in, say, Rindler sector $I$.  Let each
oscillator be a charge bound harmonically to an ion of an accelerated
dielectric medium. Each oscillator amplitude produces an electric
polarization in the dielectric medium. The manner in which these
oscillators respond collectively to an electromagnetic field is
characterized by the electric susceptibility and hence by the
refractive index of the accelerated medium.  The Maxwell field
equations, with its displacement field related to its electric field
by this refractive index, give an exact long wave description of the
interaction between the coherent motion of the oscillators and the
electromagnetic field.

Next consider a monochromatic (i.e. of Rindler frequency $\omega$)
electromagnetic wave entering Rindler sector $I$ from the past event
horizon ($V=0,U<0$), as shown in Figures \ref{fig:reflected modes},
\ref{fig:WKB modes}, or \ref{fig:reflected phases} for example.  In
the process of being reflected by the pseudo-gravitational potential,
this wave ($\propto {\mathcal{A}} \exp [-i{\mathcal{S}}^U] \sim
e^{-i(\omega \tau -ln \xi)}$) first goes through the accelerated
dielectric one way, and then, after reflection, this wave comes back
($\propto {\mathcal{A}} \exp [i{\mathcal{S}}^V] \sim e^{-i(\omega \tau
  +ln \xi)}$) the other way. In so doing the wave acquires a
well-defined phase shift.  Its magnitude, $2\delta_I(\omega)$ (not to
be confused with the interference parameters $\delta$ of
Eqs.(\ref{eq:theta constructive interference parameter}) and
(\ref{eq:constructive interference parameter}) ), depends on the
thickness of the dielectric as well as on the Rindler frequency of the
wave. The phase shifted wave escapes from Rindler sector $I$ through
its event horizon. Upon entering Rindler sector $F$, the wave combines
with a wave which suffered an analogous phase shift,
$\delta_{II}(\omega)$, in Rindler sector $II$. Thus the particle
amplitude, which in $P$ is
\[
B^+_\omega =\displaystyle
e^{-\pi \omega}{\mathcal{A}} e^{\displaystyle i{\mathcal{S}}^V}
             +{\mathcal{A}} e^{\displaystyle  -i{\mathcal{S}}^U}~~,
\]
becomes a phase-altered amplitude which in $F$ is equal to
\begin{eqnarray}
 & e^{-\pi \omega}&{\mathcal{A}} e^{\displaystyle  i{\mathcal{S}}^U} 
e^{i2\delta_{II}}
          +{\mathcal{A}} e^{\displaystyle -i{\mathcal{S}}^V}e^{i2\delta_{I}}\\
&=&
\left\{ i
\frac{\sin (\delta_{II}-\delta_I)}{\sinh \pi\omega} e^{i(\delta_{II}+ \delta_{I})}\right\}B^-_\omega 
+\left\{\frac{e^{\pi\omega}e^{i2\delta_{I}} -e^{-\pi\omega}e^{i2\delta_{II}}}{\sinh \pi\omega}\right\}
B^+_\omega 
\end{eqnarray}
The coefficient of the mode $B^-_\omega$ has become nonzero. This shows 
that the antiparticle amplitude expressed by the first term
\be
\Delta {\mathcal A}(\omega)=
i \frac{\sin (\delta_{II}-\delta_I)}{\sinh \pi\omega} 
e^{i(\delta_{II}+ \delta_{I})}
\label{eq:antiparticle amplitude}
\ee
depends on the relative phase $\delta_{II}(\omega)-\delta_I(\omega)$. 
The antiparticle
amplitude oscillates as a function of this relative phase difference.
This oscillatory behavior is the interference between the two partial
waves emerging from $I$ and $II$.

The antiparticle amplitude, and hence the amplified mode (creation of
quanta), observed in $F$ is therefore due to the difference in the
refractive indices of the dielectric media in $I$ and $II$.

\section{CONCLUSION}

The best way of summarizing this article is by starting to compare the motion
of a particle as governed by classical mechanics via
\begin{equation}
\frac{d^2x^\mu}{ds^2}~+~\Gamma^\mu_{\alpha\beta}\frac{dx^\alpha}{ds}
\frac{dx^\alpha}{ds}=0~~,
\label{eq:geodesic}
\end{equation}
or equivalently \cite{Power and Wheeler},\cite{Gerlach1969} by
\[
\frac{\partial S}{\partial x^\alpha}\frac{\partial S}{\partial x^\beta}
g^{\alpha\beta}~+~m^2=0~~~\textrm{plus}~~~
\left(
\begin{array}{c}
\textrm{``the~principle~of}\\
\textrm{constructive~interference''}
\end{array}\right)~~,
\]
with the motion of a particle as governed by wave mechanics via
\[
{\partial^2 \psi \over \partial t^2}-
{\partial^2 \psi \over \partial z^2}+ (k^2_x +k^2_y +m^2) \psi =0.
\]
One recalls that the classical mechanical
description of the free-float (``inertial'') motion of spinless
particles is the high mass limit obtained from a very special set of
wave packet solutions to the Klein-Gordon equation. Each wave packet must
contract and expand at an asymptotically non-relativistic rate, i.e. its
explosivity index $\varepsilon$ must be much smaller than one (Sections VII.B
and VII.C.1). In that case each wave packet is characterized by its proper
Minkowski frequency, which in quantum mechanics is the particle's
Compton frequency.
Increasing this frequency results in (a) the reduction of the
particle's wave packet to the location of the wave packet crest obtained from
the principle of constructive interference, (b) the reduction of the world tube
of the wave packet to the razor-sharp history of a classical particle,
and thus (c) the assignment of a needle-sharp tangent vector to each point event
on this razor-sharp world line.

The identification of the explosivity index ( $\varepsilon$ ) of a complete
set of Klein-Gordon wave packet histories permits us to grasp a new aspect
of relativistic wave mechanics:

\noindent Consider relativistically expanding wave packets which 
are characterized by a common explosivity index which is very large.
Each of their evolutions is an interference process which takes place
in the nature-given Lorentzian Mach-Zehnder interferometer of the four
Rindler sectors $P,I,II,$ and $F$.  This process, nature's spacetime
version of the double slit interference process, lends itself to a
simple mathematical description (Section VII), and is summarized by
\[
P\rightarrow (I,II) \rightarrow F~~,
\]
i.e. amplitude splitting in $P$ (Figure \ref{fig:amplitude
  splitting}), reflections in $I$ and $II$ (Figures \ref{fig:reflected
  modes} and \ref{fig:reflected phases}), and amplitude recombination,
i.e. interference, in $F$ (Figures \ref{fig:WKB modes} and
\ref{fig:reflected phases}).

Roughly speaking, this process is the Fourier dual (Section VII) of
the familiar process of a low-$\varepsilon$ wave packet contracting
and re-expanding slowly as it traces out its history in an inertial
frame of reference. In the limit of large mass, one recovers the
classical mechanics, Eq.(\ref{eq:geodesic}), of a free particle from
the \emph{low}-$\varepsilon$ and hence inertial-frame-observed wave
packet history.  What is the nature and the utility of the large mass
limit of the \emph{high}-$\varepsilon$ and hence
Rindler-frame-observed wave packet histories?  Space limitations
demand that we consign the answer to this question to a separate
paper.

Our path through the wave mechanical landscape leading to nature's
spacetime interferometer has been paved with the
\emph{Minkowski-Bessel modes} and the \emph{Klein-Gordon orthonormal
  wave packets} constructed from them.  The key markers along this
path have been the \emph{phase space} induced by these wave packets
and the \emph{Lorentz version of the Mach-Zehnder interferometer}
induced by the Minkowski-Bessel modes.  Travelling along this path, we
got a better understanding of the physical role of finite-time
detectors, and caught a glimpse of 
amplification of radiation scattered by an inhomogeneous accelerated 
dielectric.
\section{ACKNOWLEDGEMENT}

The author appreciates the pleasant discussions, which were both
stimulating and valuable, with Vladimir Belinski, Nikolai Narozhny, and
L. Sriramkumar during the 8th Marcel Grossmann Meeting in Jerusalem.
The author also would like to thank Derek Gerlach for several
valuable remarks and making available his MATLAB expertise.

\end{document}